\documentclass[prd,preprint,showpacs,superscriptaddress,amssymb,floatfix]{revtex4}
\usepackage{epsfig}
\usepackage{latexsym}
\usepackage{graphicx}
\usepackage{hyperref}
\usepackage{bm}
\usepackage{rotating}

\begin{document}

\newcommand\cs{}

\bibliographystyle{apsrev}

\preprint{CU-TP-1160, Edinburgh 2007/3, RBRC-649, KEK-TH-1138, BNL-NT 06/29}

\title{\Large\bf Localization and chiral symmetry in 2+1 flavor domain wall QCD}

\author{David J.~Antonio}
\affiliation{SUPA, School of Physics, The University of Edinburgh, Edinburgh EH9 3JZ, UK}

\author{Kenneth C.~Bowler}
\affiliation{SUPA, School of Physics, The University of Edinburgh, Edinburgh EH9 3JZ, UK}

\author{Peter A.~Boyle}
\affiliation{SUPA, School of Physics, The University of Edinburgh, Edinburgh EH9 3JZ, UK}

\author{Norman H.~Christ}
\affiliation{Physics Department, Columbia University, New York, NY 10027, USA}

\author{Michael A.~Clark}
\affiliation{Center for Computational Science, Boston University, 3 Cummington St, Boston, MA 02215, USA}

\author{Saul D. Cohen}
\affiliation{Physics Department, Columbia University, New York, NY 10027, USA}

\author{Chris Dawson}
\affiliation{RIKEN-BNL Research Center, Brookhaven National Laboratory, Upton, NY 11973, USA}

\author{Alistair Hart}
\affiliation{SUPA, School of Physics, The University of Edinburgh, Edinburgh EH9 3JZ, UK}

\author{Balint Jo\'o}
\affiliation{SUPA, School of Physics, The University of Edinburgh, Edinburgh EH9 3JZ, UK}
\affiliation{Jefferson Laboratory MS 12H2, 12000 Jefferson Avenue, Newport News, VA 23606, USA}

\author{Chulwoo Jung}
\affiliation{Physics Department, Brookhaven National Laboratory, Upton, NY 11973, USA}

\author{Richard D.~Kenway}
\affiliation{SUPA, School of Physics, The University of Edinburgh, Edinburgh EH9 3JZ, UK}

\author{Shu Li}
\affiliation{Physics Department, Columbia University, New York, NY 10027, USA}

\author{Meifeng Lin}
\affiliation{Physics Department, Columbia University, New York, NY 10027, USA}

\author{Robert D.~Mawhinney}
\affiliation{Physics Department, Columbia University, New York, NY 10027, USA}

\author{Christopher M.~Maynard}
\affiliation{EPCC, School of Physics, The University of Edinburgh, Edinburgh EH9 3JZ, UK}

\author{Shigemi Ohta}
\affiliation{Institute of Particle and Nuclear Studies, KEK,
Tsukuba, 305-0801, Japan}
\affiliation{Department of Physics,
Sokendai, Tsukuba, 305-0801, Japan}
\affiliation{RIKEN-BNL Research Center, Brookhaven National Laboratory, Upton, NY 11973, USA}

\author{Robert J.~Tweedie}
\affiliation{SUPA, School of Physics, The University of Edinburgh, Edinburgh EH9 3JZ, UK}

\author{Azusa~Yamaguchi}
\affiliation{SUPA, Department of Physics \& Astronomy, University of Glasgow, Glasgow G12 8QQ, UK}

\collaboration{RBC and UKQCD Collaborations : CU-TP-1160, Edinburgh 2007/3, RBRC-649, KEK-TH-1138, BNL-NT 06/29}


\date{\today}

\begin{abstract}
We present results for the dependence of the residual mass of domain wall
fermions (DWF) on the size of 
the fifth dimension and its relation to the 
density and localization properties of low-lying eigenvectors of the 
corresponding hermitian Wilson Dirac operator relevant to simulations of 2+1 
flavor domain wall QCD. Using the DBW2 and Iwasaki gauge actions, we generate ensembles of 
configurations with a $16^3\times 32$ space-time volume and an extent of 8 
in the fifth dimension for the sea quarks.  We demonstrate the existence 
of a regime where the degree of locality, the size of chiral symmetry 
breaking and the rate of topology change can be acceptable for inverse 
lattice spacings $a^{-1} \ge 1.6$ GeV.
\end{abstract}

\maketitle

\newpage

\section{Introduction}

In this paper we present a study of three closely related quantities 
that are important for the successful application of the domain wall
fermion (DWF) formulation to full QCD calculations with 2+1 flavors of light
quarks.  The first of these is the size of the explicit chiral symmetry 
breaking that occurs for domain wall fermions due to the
finite extent of the lattice in the fifth dimension.  
This is usually characterized by the
residual mass, $m_{\rm res}$, whose dependence on the gauge action,
gauge coupling and the extent of the fifth dimension is discussed.

The second topic covers the spectral properties of the four-dimensional
hermitian Wilson Dirac operator evaluated at large, negative mass, $-M_5$.
All current formulations of lattice chiral symmetry make use of a negative--mass 
(hermitian) Wilson Dirac operator, $H_W = \gamma_5 D_W$, and 
intellectually derive from the Kaplan approach \cite{Kaplan:1992bt},
whether by taking an analytic limit as in the overlap operator
\cite{Neuberger:1997fp}, or by directly simulating a finite fifth 
dimension \cite{Furman:1994ky} of size $L_s$.  
Auxiliary information on the nature of the 
spectrum of $H_W$ is required to have confidence that a simulation 
at fixed lattice spacing lies in the universality class of QCD.  
This topic is closely connected to the properties of the Aoki phase 
of the related operator $D_W$ \cite{Aoki:1983qi}.  If a chiral fermion calculation 
is attempted for a gauge action and a negative fermion mass that is too 
close to the Aoki phase, undesirable non-locality or enhanced explicit 
chiral symmetry breaking may result.  Thus, we will examine 
both the density of the low-lying eigenmodes of $H_W$ and their 
locality properties.

Finally, we examine the ergodicity of the sampling of topological 
charge that results for these choices of gauge action and negative 
Wilson Dirac mass using the rational hybrid Monte Carlo method of Clark 
and Kennedy \cite{Clark:2003na,Clark:2004cp}.  Since a change in 
the topological charge of an evolving lattice configuration is 
expected to be accompanied by a zero mode of $H_W$, a choice of the 
gauge action which suppresses such zero modes will inhibit topology 
change, possibly rendering the Monte Carlo sampling
non-ergodic in practice.

Based on these studies, we conclude that lattice QCD calculations
with 2+1 flavors of light quarks are possible with adequate locality,
small explicit chiral symmetry breaking and adequate sampling of 
topological charge for inverse lattice spacings of 1.6 GeV or larger.
In particular, the rapid decrease in the 
rate of change of topological charge for
the DBW2 action with decreasing lattice spacing favours the choice 
of Iwasaki action---a choice that has been adopted for the large-scale 
ensemble generation of the RBC and UKQCD collaborations using the 
QCDOC computers\cite{Boyle:2001dc,Boyle:2003mj,Boyle:2005gf} at the RIKEN-BNL Research Center and the University 
of Edinburgh.

The structure of this paper is as follows.  
In Section~\ref{secAoki} we describe the Aoki phase of the Wilson 
Dirac operator in light of recent research and the connection to the 
spectrum and localization of eigenmodes of $H_W$.
In Section~\ref{secEigenmodesIntro} we review the importance of 
the spectrum of $H_W$ for 
lattice formulations that accurately realize chiral symmetry.  
The explicit breaking of chiral symmetry that results from finite 
$L_s$ for domain wall fermions is discussed in Section~\ref{secMres} 
and the expected dependence of the residual mass on $L_s$ motivated.
In Section~\ref{secAction} we describe the details of the 
gauge and DWF actions used in our simulations, while in 
Section~\ref{secEigenmodes} we present a microscopic study 
of the low-lying eigenvalues and eigenmodes of $H_W$
on our ensembles.  In Section~\ref{secMresResults} 
we present a study of the dependence on $L_s$ of the residual 
mass for valence quarks on each of the (fixed sea quark $L_s$) 
ensembles. The $L_s$ dependence of these results are fitted to 
the form motivated in Section~\ref{secMres} and conclusions about the 
low-lying spectrum of the transfer matrix in the fifth dimension are
drawn and related to our results from Section~\ref{secEigenmodes}.  
Finally, in Section~\ref{secTopology}, we examine the dependence 
of the topological charge with evolving Monte Carlo time and 
demonstrate the correlation between reduced DWF chiral symmetry 
breaking and a suppression in the rate of change of topological 
charge.   Appendix~\ref{sec:transfer_matrix} summarizes the 
transfer matrix formalism of Furman and Shamir which is used 
extensively in the residual mass discussion of Section~\ref{secMres}.

\section{Aoki phase diagram of lattice QCD}
\label{secAoki}

The properties of lattice chiral
fermions are intimately connected with the properties of the hermitian
Wilson Dirac operator $H_W$ for negative mass $-M_5$.  Of specific interest
is the ``super-critical'' region, $-8 \le -M_5 \le 0$, in which both $H_W$ and $D_W$
can have zero eigenvalues.  This region shows a surprisingly 
rich and interesting phase structure, first conjectured and described by Aoki 
\cite{Aoki:1983qi,Aoki:1985mk,Aoki:1986xr,Aoki:2001su} in both the quenched 
and dynamical cases \footnote{While this paper addresses ``non-quenched''
physics with 2+1 dynamical flavors, the dynamical quark masses 
applied to surface states in our simulations
are not related to the value of $M_5$ used in the domain wall construction
so it is the properties of the quenched Aoki phase which are most relevant
for our calculations.}.

The Aoki phase was originally defined as a region in which a non-vanishing
pionic condensate spontaneously breaks flavor and parity for the two-flavor 
dynamical case, with an associated flavor non-singlet Goldstone pion.
In the quenched case, if one examines Green's functions made from a single
flavor of quark, (discrete) parity is spontaneously broken.   The
propagator for a flavor-singlet pion describes a massless state
at the critical line separating the normal from the Aoki phase, but
which becomes massive within the interior of the Aoki phase.  For Green's 
functions containing two quark flavors, connected by a vanishing 
flavor-breaking mass term, parity and flavor are broken by the pionic 
condensate $\overline{\psi}\gamma_5\tau^a \psi$ and massless, flavored Goldstone modes 
exist inside the Aoki phase.

Since the hermitian Wilson Dirac operator obeys a Banks-Casher relation, a 
non-zero density of near-zero modes is associated with this condensate in both 
the quenched and dynamical cases. This non-zero density is a strong coupling 
effect, arising from the disorder in the gauge fields which 
characterizes the strong coupling limit.  As the coupling becomes weak, the 
gauge field becomes more ordered and such a density of near-zero modes is 
expected to disappear, except for values of $M_5$ where the 
free Dirac operator 
itself has zero modes, $-M_5 = 0, -2, -4, -6$ and $-8$.

However, the picture suggested by the simple summary above is not complete.  
It was observed \cite{Aoki:2001su,Aoki:2001de} that the interior of the quenched Aoki 
phase consists of two qualitatively different regions bounded by a critical 
line $g_c(-M_5)$.  Above $g_c(-M_5)$ near-zero modes exist which are delocalized 
while below $g_c(-M_5)$, only localized near-zero modes appear.  In both 
regions the pionic condensate and density of near-zero modes are non-zero, 
but their localization properties differ fundamentally.

This picture was substantially refined and solidified by Golterman et al.
\cite{Golterman:2003qe,Golterman:2004cy,Golterman:2005fe,Svetitsky:2005qa}
who, in close analogy with consequences of randomness in condensed matter
physics, introduced to QCD the concept of a non-zero mobility edge $\lambda_c$,
as the critical eigenvalue of $H_W$ above which eigenstates are extended 
and below which all eigenstates are localized.  They also applied the McKane and 
Stone localization escape from Goldstone's theorem to lattice QCD at non-zero 
lattice spacing in the quenched case.  Specifically, they showed that two 
quenched flavors of Wilson fermion with a non-zero mobility edge can display a spontaneous 
breaking of a continuous symmetry \emph{without} a corresponding Goldstone 
boson.  This observation is key to the correct functioning of all the 
various formulations of Kaplan fermions wherever the kernel $H_W$ has a 
non-zero density of low modes.  

\begin{figure}[hbt]
\includegraphics[width=1.0\textwidth]{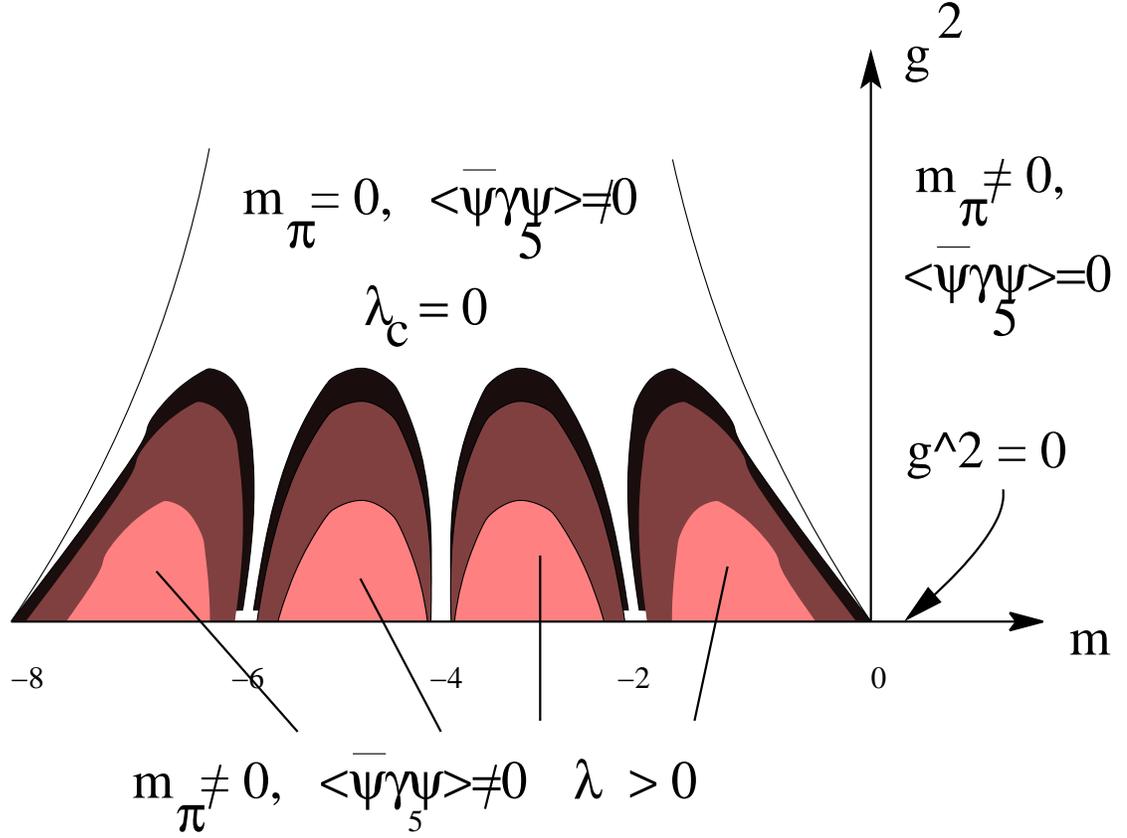}
\caption{\label{aokiphase}
Schematic diagram of the quenched Aoki phase. A pionic condensate and corresponding
non-zero density of near zero modes is developed throughout most of
the negative mass region. In the coloured sectors, a non-zero mobility edge is
developed and the contours could equally well represent either decreasing low mode density,
decreasing pionic condensate or increasing $\lambda_c$ as one moves towards the continuum
limit at $g^2=0$.
}
\end{figure}

This qualitative picture is displayed in Figure~\ref{aokiphase}.  Following
conventional terminology, we refer to the Aoki phase as that region of this 
diagram in which $\langle\overline\psi\gamma^5\tau^a\psi\rangle \ne 0$ and 
$g^2 > g^2_c(-M_5)$.  In this region the coupling is sufficiently large that 
the mobility edge is zero and long-range correlations result from massless 
flavor-non-singlet pions.  The near-zero modes of 
$H_W$ are delocalized.  This is a dangerous region for chiral fermions,
 with lattice artifacts producing 
unphysical, long-distance correlations.  For weaker coupling 
$g^2 \le g^2_c(-M_5)$, these long-distance correlations have disappeared.
The Wilson Dirac eigenvectors with low 
eigenvalues are localized on the lattice scale
and all delocalized modes have eigenvalues which, if expressed in 
physical units, are at least as large as $\lambda_c(-M_5)/a$.  These delocalized low
modes do not introduce unphysical non-locality into the corresponding 
formulation of lattice chiral fermions but, as discussed in the next
section, they enhance the explicit violation of chiral symmetry for 
domain wall fermions with finite $L_s$.

\section{Low-lying spectrum of $H_W$ and lattice chiral fermions}
\label{secEigenmodesIntro}

We now discuss in more detail the relation between the spectrum of the
hermitian Wilson Dirac operator $H_W$ and the properties of domain wall fermions.
Let us first address the issue of locality.  While the domain wall
Dirac operator contains only nearest-neighbor couplings in five dimensions 
(i.e. is ultra-local), the four dimensional effective
low energy theory which it produces may contain unphysical non-locality.
One approach to 
investigating this question is to examine the locality of the overlap 
operator that results in the $L_s \rightarrow \infty$ limit
\cite{Neuberger:1997bg,Kikukawa:1999sy}. 
This was 
done by Hernandez et al. \cite{Hernandez:1998et}
using $H_W$ as the kernel in the sign function of the overlap operator. They showed
that a gap in the spectrum of the eigenstates of $H_W $ guarantees locality of the 
resulting overlap operator. The proof relies on the ultra-local nature of $H_W$ implying
ultra-locality of a finite polynomial of $H_W$.
The ``Shamir'' kernel $K_S$ for the overlap operator
that corresponds to the
$L_s=\infty$ limit of domain wall fermions 
can be taken as $K_S = H_W (2+D_W)^{-1}$ 
\cite{Borici:1999zw,Borici:1999da,Edwards:2000qv}
and it is not manifestly ultra-local. 
However, the operator $2+D_W$ in the denominator of $K_S$ is the Wilson
Dirac operator with a $2-M_5$ mass term and hence, provided $M_5 < 2$ is 
outside the super-critical region
ensuring that this kernel is exponentially localized. 
The Hernandez et al. proof is therefore
applicable to domain wall fermions under the modification that a finite polynomial of an exponentially
local operator is also exponentially local.

However, as discussed above, the assumed gap in the spectrum of $H_W$ 
is neither expected nor observed in lattice QCD, at least for the range
of couplings in which calculations are normally performed.  Fortunately, 
as argued by Golterman et al.~\cite{Golterman:2003qe}, the 
detailed properties of the Wilson Dirac operator in the super-critical region 
described above are sufficient to imply that the overlap operator 
is appropriately local (provided we are outside of the Aoki phase and
$\lambda_c(-M_5,g) > 0$).  If, as is the case in this picture, one assumes
that the spectrum of extended states shows a gap (the region 
$0 \le \lambda \le \lambda_c$) so that any eigenstates within the gap
are localized, then the resulting overlap operator will also be 
localized (even though the density of low modes is non-zero).
%
This is supported by numerical evidence at currently affordable couplings.

Finally, we examine a second difficulty faced by domain wall fermions and 
related approaches, which is closely connected with these low-lying 
modes of $H_W$. The residual chiral symmetry breaking effects seen 
at finite $L_s$ for domain wall fermions depend in detail on the 
densities and sizes of the modes of the hermitian matrix $H_T$.
This is reviewed in Appendix~\ref{sec:transfer_matrix}, and $H_T$ is defined 
in Eq.~\ref{eq:H_def}.   $H_T$ is used to construct the 
Fock-space transfer matrix in the 
fifth dimension,  $T$ defined in Eq.~\ref{eq:T_def}.  
This transfer matrix describes
explicitly how the left and right walls are coupled when $L_s$ is
finite. While the relation between the operators $H_W$ and $H_T$
is somewhat complex, it can be shown that their zero modes
coincide~\cite{Furman:1994ky}, and also
that $ -\log T \equiv H_T = 2 \tanh^{-1} K_S$, where $K_S = \frac{H_W}{2+D_W}$
\cite{Borici:1999zw,Borici:1999da,Edwards:2000qv}.
Further, the corresponding approximation to the overlap operator, and its
$L_s\to\infty$ limit are:
\cite{Neuberger:1997bg}

\begin{equation}
\label{eq:Doverlap}
\begin{array}{cccc}
   & D_{\rm ov}^{\rm approx}  & = & \frac{1}{2}\left[ 1+m + (1-m)\gamma_5 \tanh ( \frac{L_s}{2} H_T ) \right]\\
\to& D_{\rm ov}               &=& \frac{1}{2}\left[ 1+m + (1-m)\gamma_5 {\rm sgn} H_T ) \right],
\end{array}
\end{equation}
or equivalently,
\begin{equation}
\begin{array}{cccc}
& D_{\rm ov}^{\rm approx} &=&
 \frac{1}{2}\left[ 1+m + (1-m)\gamma_5 \tanh ( L_s \tanh^{-1} K_S) 
       \right]\\
\to & D_{\rm ov}& = &\frac{1}{2}\left[ 1+m + (1-m)\gamma_5 {\rm sgn} K_S ) \right].
\end{array}
\end{equation}
The equality of the operators ${\rm sgn} H_T$ and ${\rm sgn} K_S$
follows easily from the relation $H_T = 2 {\rm tanh}^{-1}K_S$ and
the recognition that the inverse hyperbolic tangent is a function
which preserves the sign of its argument.

Thus, the near-zero modes of $H_W$ discussed above are expected to 
correspond to near-zero modes of $H_T$, modes
which (in the absence of an explicit mass term) will dominate the coupling of
the left- and right-handed sectors of low-energy domain wall QCD.
Similarly, the mobility edge structure of $H_W$ is also expected to describe $H_T$.
As is exploited in the next section, this implies a simple
structure for the asymptotic $L_s$-dependence of the residual mass.
The near-zero localized modes will give a power behavior ($1/L_s$),
while the extended modes above the mobility edge give an exponential
fall-off ($e^{-\lambda_c L_s}/L_s$).  This dependence will be used 
as one of several diagnostics to investigate the nature of the 
$H_T$ spectrum in our simulations.

The strong-coupling behaviour of the Wilson Dirac operator does not 
leave one at liberty to simulate QCD at arbitrarily coarse lattice 
spacings.  For a given lattice action, the phase boundary where the 
gap in the spectrum of extended states vanishes defines the coarsest 
lattice spacing at which the formulation makes sense and implies a 
minimum cost that must be paid to have the formulation under control. 
The phase boundary is action dependent and, in practice, we will wish 
to stay well within the phase such that the mobility edge is not small.

We have performed the first numerical study of the localization 
of the hermitian Wilson Dirac operator with three mass degenerate
flavors of light \emph{dynamical} quarks and demonstrate that a programme
of 2+1 flavor dynamical DWF simulations of QCD is rendered affordable 
by our current QCDOC computer systems.  These considerations are directly 
relevant for all chirally symmetric lattice QCD formulations,
and we shall later illustrate the smooth 
connection between finite $L_s$ and the $L_s\to\infty$ overlap limit 
with numerical data for the plaquette. Our work
has been presented at Lattice 2005 \cite{Antonio:2005wm}, and later,
while this paper was being completed, a broadly similar study was presented
at Lattice 2006 using dynamical overlap fermions~\cite{Yamada:2006fr}.

\section{Residual mass}
\label{secMres}

In this section we discuss the residual mass in some detail, including a
careful discussion of its dependence on $L_s$ as predicted by the transfer
matrix analysis of Furman and Shamir.  As we will demonstrate and has
been worked out previously \cite{Golterman:2003qe,Golterman:2004cy,
Golterman:2005fe,Svetitsky:2005qa,Christ:2005xh,Antonio:2005wm}, this
dependence on $L_s$ permits the effects of both the localized near-zero
modes and the extended modes above the mobility edge to be recognized.
This gives useful information about the general character of chiral 
symmetry breaking as well as an understanding of the origin of the 
residual mass itself.

In the limit of small lattice spacing, or equivalently weak gauge
coupling, the spectrum of the domain wall Dirac operator for a typical gauge 
background is expected to have a physical, four-dimensional component 
with eigenvalues $\lambda \sim \Lambda_{\rm QCD}$ as well as unphysical, 
five-dimensional states with $\lambda \sim 1/a$.  The low-lying modes
permit the accurate simulation of QCD while the large eigenvalues are
lattice artifacts similar to, but more numerous than, the unphysical
large eigenvalues found in other lattice fermion formulations.
Ideally, the physical four-dimensional states will be bound near 
the $s=0$ and $s=L_s-1$ walls.  To the extent that $L_s$ is large, 
there should be little mixing between the left-handed chiral states 
localized on the left ($s=0$) wall and the right-handed chiral states
localized on the right ($s=L_s-1$) wall.

The low-energy properties of this approximately chiral theory can be
described by an effective Lagrangian, ${\cal L}_{\rm eff}$.  If effects 
coming from non-zero lattice spacing or arising from the overlap between 
the left- and right-handed states are neglected, this effective Lagrangian 
will be precisely that of QCD.  The corrections coming from these effects
can be described by adding extra operators to ${\cal L}_{\rm eff}$.  The
only relevant operator of this sort is a dimension-3 mass term.
The next most important term is the familiar dimension-5 Sheikholeslami-Wohlert 
term.  Thus,
\begin{equation}
{\cal L}_{\rm eff} = {\cal L}_{\rm QCD} + m_{\rm res}\overline q q 
                           + c_5 \overline q \sigma^{\mu\nu} F^{\mu\nu} q + \ldots
\label{eq:eff_L}
\end{equation}
In this paper we focus on the coefficient, $m_{\rm res}$, of this
residual mass term as an important measure of the violation of chiral
symmetry arising from the finite extent ($L_s$) in the fifth dimension.
In this section, we will discuss the dependence of $m_{\rm res}$ on 
$L_s$ by using the transfer matrix formalism of Furman and 
Shamir~\cite{Furman:1994ky}, described in 
Appendix~\ref{sec:transfer_matrix} which is a prerequisite for 
those unfamiliar with this formalism.  

\subsection{Transfer matrix in the fifth dimension}

Using the transfer matrix approach, the approximate chiral symmetry of 
domain wall fermions is realized by writing a general Green's function by
semi-independent products of left- and right-handed operators.   As 
is described in Appendix~\ref{sec:transfer_matrix}, if we arrange the 
field variables appearing in a general Green's function by segregating the 
left- and right-handed fields into two products, ${\cal O}_L[P_L q, \overline q P_R]$
and ${\cal O}_R[P_R q, \overline q P_L]$ then, we can write
\begin{eqnarray}
\left\langle {\cal O}_L[P_L q,\overline q P_R]\; 
              {\cal O}_R[P_R q,\overline q P_L]\right\rangle_{L_s}\hskip -0.5in && 
\label{eq:Gmann2op} \\
&&=Z(L_s) \; {\rm tr}\left\{ {\cal O}_L[P_L\hat a, -\hat a^\dagger P_R] 
                 \hat T^{L_s}{\cal O}_R[P_R \hat a, \hat a^\dagger P_L] {\cal{O}}(m_f)\right\}.
\nonumber
\end{eqnarray}
The left-hand side of this equation is a general, DWF Green's function 
defined as a Grassmann integral over the 5-dimensional domain wall 
field variables $\overline{\Psi}(x,s)$ and $\Psi(x,s)$ using the usual 
DWF action, $S_F = -\overline{\Psi} D^{\rm DWF}\Psi$.  Here the Grassmann 
fields $\overline{\Psi}(x,s)$ and $\Psi(x,s)$ are functions of a space-time 
coordinate $x$ and fifth-dimensional coordinate $s$, $0 \le s \le L_s-1$.  
The domain wall Dirac operator $D^{\rm DWF}$ is that of 
Shamir \cite{Shamir:1993zy}, and Furman and Shamir \cite{Furman:1994ky}
and in our notation is given by
\begin{equation}
  D^{\rm DWF}_{x,s; x^\prime, s^\prime}(M_5, m_f)
   = \delta_{s,s^\prime} D^\parallel_{x,x^\prime}(M_5)
   + \delta_{x,x^\prime} D^\bot_{s,s^\prime}(m_f)
\label{eq:D_dwf}
\end{equation}
\begin{eqnarray}
D^\parallel_{x,x^\prime}(M_5) & =&
  {1\over 2} \sum_{\mu=1}^4 \left[ (1-\gamma_\mu)
  U_{x,\mu} \delta_{x+\hat\mu,x^\prime} + (1+\gamma_\mu)
  U^\dagger_{x^\prime, \mu} \delta_{x-\hat\mu,x^\prime} \right] 
  \nonumber \\
  & + & (M_5 - 4)\delta_{x,x^\prime}
\label{eq:D_parallel}
\end{eqnarray}
\begin{eqnarray}
D^\bot_{s,s^\prime}(m_f) 
     &=& {1\over 2}\Big[(1-\gamma_5)\delta_{s+1,s^\prime} 
                 + (1+\gamma_5)\delta_{s-1,s^\prime} 
                 - 2\delta_{s,s^\prime}\Big] \nonumber\\
     &-& {m_f\over 2}\Big[(1-\gamma_5) \delta_{s, L_s-1}
       \delta_{0, s^\prime}
      +  (1+\gamma_5)\delta_{s,0}\delta_{L_s-1,s^\prime}\Big].
\label{eq:D_perp}
\end{eqnarray}
The physical four-dimensional Grassmann fields, $q(x)$ and $\overline q(x)$,
are constructed from the five-dimensional fields $\overline{\Psi}(x,s)$ and 
$\Psi(x,s)$ according to Eqs.~\ref{eq:q_def} and \ref{eq:qbar_def}.

The right-hand side of Eq.~\ref{eq:Gmann2op} is the trace of a product of
operators acting on a many-particle Fock space.  In contrast to the usual
field theory setting, these particles are located in a four-dimensional 
coordinate space.  Thus, the fields $a_x$ and $a_x^\dagger$ depend on the 
space-time position $x$, as well as other spin and color indices which 
are not shown.   These operators, as well as the transfer matrix $\hat T$,
and mass operator ${\cal{O}}(m_f)$, are defined in Appendix~\ref{sec:transfer_matrix}.  
The factor $Z(L_s)$ is an $L_s$-dependent normalization factor.  The
mass $M_5$ is the domain wall height.  The mass $m_f$ is the bare mass of the physical fermions.

In the limit in which $m_f=0$ and $L_s\to\infty$, $\hat{T}^{L_s}$ is viewed as projecting onto 
the state with the largest eigenvalue (normalized to be unity), and the 
right-hand side of Eq.~\ref{eq:Gmann2op} becomes the product of two 
completely independent matrix elements corresponding to isolated left- 
and right-handed theories (see Eq.~\ref{eq:chiral_factorization}).  
This exact chiral symmetry is violated by the operator ${\cal{O}}(m_f)$ 
which represents the explicit mass term and mixes the left- and 
right-handed factors.  Similarly, the corrections to the asymptotic 
limit $\hat T^{L_s} \to |0_H\rangle \langle 0_H|$ of Eq.~\ref{eq:T_corr} 
introduce chiral symmetry breaking arising from finite $L_s$.

In the above discussion, and that to follow, we are ignoring the 
effects of anomalous chiral symmetry breaking.  Such effects 
require modifications to our discussion for background gauge
fields with non-zero Pontryagin index.  For such gauge fields
the state $|0_H\rangle$ will carrying a fermion number different
from that of $|0_S\rangle$, thereby introducing flavor-singlet
chirality correlations between the otherwise independent left
and right-hand factors in the first term on the right-hand
side of Eq.~\ref{eq:Gmann2op} \cite{Narayanan:1995gw}.  In the 
interests of simplicity, we do not consider such gauge configurations 
here.  However, the effects of these configurations do not alter 
our conclusions.

\subsection{Residual mass and the transfer matrix}

We will now explicitly study the first corrections to the large $L_s$ limit 
of $\hat T^{L_s}$ in Eq.~\ref{eq:Gmann2op} where, for clarity, we set $m_f=0$ 
and use the relation ${\cal{O}}(m_f=0) = |0_S\rangle\langle 0_S|$ discussed in 
Appendix~\ref{sec:transfer_matrix}:
\begin{eqnarray}
\left\langle {\cal O}_L[P_L q,\overline q P_R]  
              {\cal O}_R[P_R q,\overline q P_L]\right\rangle_{L_s} &=& 
Z(L_s) \; \langle 0_S|{\cal O}_L[P_L\hat a, -\hat a^\dagger P_R] \label{eq:Gmann2op_2} \\
&& \hskip -1.3 in \Bigl\{|0_H\rangle \langle 0_H| 
+\sum_{k^+=1}^{N^+} e^{-E^+_{k^+}L_s}\hat o_{k^+}^\dagger |0_H\rangle \langle 0_H|\hat o_{k^+} 
        +\sum_{k^-=1}^{N^-} e^{-E^-_{k^-}L_s}\hat p_{k^-}^\dagger |0_H\rangle \langle 0_H|\hat p_{k^-}\Bigr\}
\nonumber \\ 
&& \hskip 1in {\cal O}_R[P_R \hat a, \hat a^\dagger P_L] |0_S\rangle.
\nonumber
\end{eqnarray}
Here the leading term, the projection operator $|0_H\rangle \langle 0_H|$,
divides the Green's function into two independent factors demonstrating the
separate flavored chiral symmetry of the left- and right-handed fermions.
The first correction permits quark number $\pm 1$ exchanges between these 
two otherwise independent sectors.  This suggests that the second and third 
terms in Eq.~\ref{eq:Gmann2op_2} should be interpreted at low energies as 
the residual mass operator $m_{\rm res} \overline q q$ expressed in this 
5-dimensional language.

A relation between this residual mass term and the second and third terms 
in the operator in curly brackets in Eq.~\ref{eq:Gmann2op_2} can be obtained
if we express the operators $\hat o_k$ and $\hat p_k$ in terms of the 
conventional 4-dimensional field $\hat a_x$, inverting Eq.~\ref{eq:a_expansion}:
\begin{eqnarray}
-\sum_n m_{\rm res}\overline q(x_n) q(x_n)
 \rightarrow && m_{\rm res}\sum_n \Bigl\{
      (\hat a^\dagger_{x_n}P_R)_\alpha|0_H\rangle\langle 0_H|(P_R \hat a_{x_n})_\alpha
\label{eq:m_res_5d} \\
      &&\hskip 0.5in  +(P_L \hat a_{x_n})_\alpha |0_H\rangle\langle 0_H|(\hat a^\dagger_{x_n}P_L)_\alpha
                   \Bigr\}  \nonumber\\
\mathrel{\mathop\approx^?}&&  
    \sum_k \sum_{n,n'} \phi^+_{k \alpha}(x_n) \phi^+_{k\beta}(y_{n'})^* e^{-L_s E_k^+}\; 
         (\hat a_{x_n}^\dagger P_R)_\alpha|0_H\rangle\langle 0_H|(P_R \hat a_{y_{n'}})_\beta
 \nonumber \\ 
+&& \sum_k \sum_{n,n'} \phi^-_{k\alpha}(x_n)^* \phi^-_{k\beta}(y_{n'}) e^{-L_s E_k^-}\; 
         (P_L\hat a_{x_n})_\alpha|0_H\rangle\langle 0_H| (\hat a_{y_{n'}}^\dagger P_L)_\beta. \nonumber
\\
\label{eq:m_res_def1}
\end{eqnarray}
Here the RHS of Eq.~\ref{eq:m_res_5d} is the usual residual mass term, $-m_{\rm res} 
\overline{q}q$, written in the transfer matrix language (see Eq.~\ref{eq:qbq2}),
while the expression in Eq.~\ref{eq:m_res_def1} is a version of 
the leading order contribution from the second and third terms in 
Eq.~\ref{eq:Gmann2op_2}.  The projection operators $P_R$ and $P_L$ can
be introduced into the right-hand side of Eq.~\ref{eq:m_res_def1} because the
other chirality will vanish in the limit $m_f=0$ \footnote{For example, 
in the left-hand factor, the wrong-chirality operators, $P_R \hat a_x$ 
and $\hat a^\dagger P_R$, do not appear in the operator expression 
${\cal O}_L[P_L q,\overline q P_R]$ and annihilate the state 
$\langle 0_S|$ and hence can be dropped for the present $m_f=0$ case.}.

We can justify this relation and estimate $m_{\rm res}$ if we assume that
the sums $\sum_k \phi^\pm_k(x_n)^\dagger \phi^\pm_k(y_{n'})$ are localized on 
the long distance scale at which $m_{\rm res}$ is defined:
\begin{equation}
K_{\alpha,\beta}(x,y) 
  = \sum_k \phi^\pm_{k,\alpha}(x)^\dagger \phi^\pm_{k,\beta}(y) e^{-L_s E_k^\pm} 
                \approx m_{\rm res} \delta^4(x-y) \delta_{\alpha,\beta}\,,
\label{eq:def_K}
\end{equation}
where $\alpha$ and $\beta$ are each spin and color indices, and we assume that 
the diagonal spin/color structure will appear and any dependence on the label 
$\pm$ will disappear when a volume and/or gauge average is performed.  
We can then approximate: $m_{\rm res} = \frac{1}{12}\int d^4 x K_{\alpha,\alpha}(x,y) 
\approx \frac{1}{12} R^4 K_{\alpha,\alpha}(y,y)$,  where the ``radius'' 
$R$ estimates the small region in $x$ that contributes and summation is performed
over the repeated spin and color index $\alpha$.  Finally, $K(y,y)$ and 
hence $m_{\rm res}$ can be determined by integrating over $y$ and using 
the orthonormality of the eigenfunctions $\phi^\pm_{k\alpha}(y)$:
\begin{eqnarray}
m_{\rm res} &=& \frac{R^4}{12} K_{\alpha,\alpha}(y,y) 
               = \frac{R^4 }{12L^3 L_t} \int d^4 y K_{\alpha,\alpha}(y,y) \\
            &=& \frac{R^4 }{12 L^3 L_t} \sum_k e^{-L_s E_k^\pm} 
               = R^4 \int_0^\infty d\lambda \rho(\lambda) e^{-L_s \lambda}
               \label{eq:spec_density} \\
            &\sim& R_e^4 \rho_e(\lambda_c) \frac{e^{-\lambda_c L_s}}{L_s} +
                             R_l^4 \rho_l(0)\frac{1}{L_s}.
           \label{eq:spec_density_2}
\end{eqnarray}
Here $\rho(\lambda)$ on the RHS of Eq.~\ref{eq:spec_density} is the density
of eigenvalues of $H_T$ per unit space-time volume, color and spin.  The final 
Eq.~\ref{eq:spec_density_2} is a generalization of Eq.~\ref{eq:spec_density}
displaying the expected contributions of extended ($e$) and localized ($l$) 
modes with possibly different radius parameters, $R_l$ and $R_e$. 
Eq.~\ref{eq:spec_density_2} should be viewed as a theoretically motivated
model for the large $L_s$ behavior of the residual mass 
where $R_l$ is the radius of the localized states with 
eigenvalue close to zero.  The radius associated with
the extended states, $R_e$, arises from the limited spatial 
resolution provided by a sum over extended states resulting
from the high-momentum cutoff intrinsic to the lattice
formulation.  This suggests $R_e \approx a$.
In this final equation we have also taken the limit of large $L_s$ to simplify the integral
over $\lambda$.  This is the ``standard'' result for the dependence of 
$m_{\rm res}$ on $L_s$, although the factor of $1/L_s$ is usually missing 
from the left-most term.  We will use Eq.~\ref{eq:spec_density_2} elsewhere 
in this paper to interprete the dependence of $m_{\rm res}$ on $L_s$.

\subsection{Residual mass and the 5-dimensional axial current}

While the above discussion provides a direct connection between corrections
to the $L_s \to \infty$ limit and the residual mass, it does not give a
simple way to compute $m_{\rm res}$.  This is typically done using a
partially conserved 5-dimensional axial current introduced by Furman
and Shamir~\cite{Furman:1994ky}:
\begin{equation}
{\cal A}^b_\mu(x) = \sum_{s=0}^{L_s-1} {\rm sign}(s -\frac{L_s-1}{2})j^b_\mu(x,s),
\end{equation}
where the current $j^b_\mu(x,s)$ is defined by
\begin{equation}
j^b_\mu(x,s) = \frac{1}{2}\left[
          \overline\Psi(x+\hat\mu,s)(1+\gamma_\mu)U^\dagger_{x+\mu,\mu} t^b \Psi(x,s)
         -\overline\Psi(x,s)(1-\gamma_\mu)U_{x,\mu} t^b \Psi(x+\mu,s)\right],
\end{equation}
and $b$ is a flavor index and $t^b$ is a generator of the flavor symmetry.  The equations 
of motion imply the conservation law
\begin{equation}
\Delta_\mu {\cal A}^b_\mu(x) = 2 m_f \overline q(x)t^b\gamma^5 q(x) + J^b_{5q}(x)
\label{eq:a_conserv}
\end{equation}
where $J^b_{5q}$ is defined in Eq.~\ref{eq:j5q_def} and
we define $(\Delta_\mu f)(x) = f(x) -f(x-\hat\mu)$.

In the limit of large $L_s$ and small lattice spacing, the extra $J^b_{5q}$
term in the conservation law in Eq.~\ref{eq:a_conserv} should describe the
chiral asymmetry of the leading dimension-3 term in the domain wall fermion
effective Lagrangian, $m_{\rm res}\overline q q$:
\begin{equation}
J^b_{5q} \approx 2m_{\rm res}\overline q t^b \gamma^5 q.
\label{eq:m_res_def2}
\end{equation}
This equation can be used directly to determine $m_{\rm res}$ by 
evaluating the ratio of the matrix elements of $J^b_{5q}$ 
and $\overline q t^b \gamma^5 q$.  In this paper, we use the ratio 
of pion correlation functions:
\begin{equation}
m_{\rm res} = \frac{\langle J^b_{5q}(x) \; \overline q t^b \gamma^5 q(y)\rangle}
      {\langle 2\overline q t^b \gamma^5 q(x) \; \overline q t^b \gamma^5 q(y)\rangle}
\label{eq:mres_axial}
\end{equation}
evaluated for $|x-y| >> a$ to determine $m_{\rm res}$ numerically.

As a consistency check, we should demonstrate that the definitions of 
$m_{\rm res}$ given in Eqs.~\ref{eq:m_res_def1} and \ref{eq:m_res_def2} 
agree.  We begin by generalizing Eq.~\ref{eq:j5q_transf} to express a 
general Green's function of $J^b_{5q}(z)$ with other physical fields,
in a form in which the latter are 
separated into the left- and right-handed factors 
${\cal O}_L[P_L q, \overline q P_R]$ and ${\cal O}_R[P_R q, \overline q P_L]$:
\begin{eqnarray}
\langle {\cal O}_L[P_L q, \overline q P_R] J^b_{5q}(z) {\cal O}_R[P_R q, \overline q P_L] \rangle && 
\nonumber \\
&& \hskip -2.0in  = Z(L_s) \langle 0_S| {\cal O}_L[P_L \hat a, -\hat a^\dagger P_R] 
                            \hat T^\frac{L_s}{2} \hat a^\dagger_z t^b\hat a_z \hat T^\frac{L_s}{2}
                             {\cal O}_R[P_R \hat a, \hat a^\dagger P_R] | 0_S\rangle.
\label{eq:j5q_me1}
\end{eqnarray}
Since we wish to compare with Eqs.~\ref{eq:m_res_5d} and \ref{eq:m_res_def1},
we can simplify this expression by integrating the position variable $z$
over space-time:
\begin{eqnarray}
 Z(L_s) \langle 0_S| {\cal O}_L[P_L \hat a, -\hat a^\dagger P_R] 
  \hat T^\frac{L_s}{2}\Bigl\{
            \sum_{k^+=1}^{N^+}\hat o^\dagger_{k^+} t^b\hat o_{k^+}
           -\sum_{k^-=1}^{N^-}\hat p^\dagger_{k^-} t^b\hat p_{k^-}\Bigr\}
  \hat T^\frac{L_s}{2}{\cal O}_R[P_R \hat a, \hat a^\dagger P_R] | 0_S\rangle.
\label{eq:j5q_me2}
\end{eqnarray}
Here we have also replaced the operators $\hat a_x$ and $\hat a_x^\dagger$ with 
those creating eigenstates of $\hat T$, using the definition in 
Eq.~\ref{eq:a_expansion} and the orthogonality relations obeyed by the 
coefficients $\phi_{k^\pm}^\pm$.  Next, we can evaluate the factors 
$\hat T^\frac{L_s}{2}$ in the limit of large $L_s$ using Eq.~\ref{eq:T_corr}.  
The leading contribution to Eq.~\ref{eq:j5q_me2} results when the next 
leading contribution to $\hat T^\frac{L_s}{2}$ (the second and third 
terms in Eq.~\ref{eq:T_corr}) is substituted for each factor:
\begin{eqnarray}
 Z(L_s) \langle 0_S| {\cal O}_L[P_L \hat a, -\hat a^\dagger P_R]
\Bigl( \sum_{k_1^+=1}^{N^+} e^{-E^+_{k_1^+}\frac{L_s}{2}}
                   \hat o_{k_1^+}^\dagger |0_H\rangle \langle 0_H|\hat o_{k_1^+} 
                       +\sum_{k_1^-=1}^{N^-} e^{-E^-_{k_1^-}\frac{L_s}{2}}
                \hat p_{k_1^-}^\dagger |0_H\rangle \langle 0_H|\hat p_{k_1^-} \Bigr) &&
\nonumber \\
&& \hskip -5.0in \Bigl\{    \sum_{k^+=1}^{N^+}\hat o^\dagger_{k^+} t^b\hat o_{k^+}
                           -\sum_{k^-=1}^{N^-}\hat p^\dagger_{k^-} t^b\hat p_{k^-}\Bigr\}
\label{eq:j5q_me3} \\
&& \hskip -6.0in \Bigl( \sum_{k_2^+=1}^{N^+} e^{-E^+_{k_2^+}\frac{L_s}{2}}
                \hat o_{k_2^+}^\dagger |0_H\rangle \langle 0_H|\hat o_{k_2^+} 
      +\sum_{k_2^-=1}^{N^-} e^{-E^-_{k_2^-}\frac{L_s}{2}}
                \hat p_{k_2^-}^\dagger |0_H\rangle \langle 0_H|\hat p_{k_2^-} \Bigr)
 {\cal O}_R[P_R \hat a, \hat a^\dagger P_L] | 0_S\rangle.
\nonumber
\end{eqnarray}
Exploiting the simple form of the operator in the curly brackets, 
Eq.~\ref{eq:j5q_me3} can be considerably simplified, yielding:
\begin{eqnarray}
 Z(L_s) \langle 0_S| {\cal O}_L[P_L \hat a, -\hat a^\dagger P_R] && \label{eq:j5q_me4} \\
&& \hskip -1.5in
\Bigl\{ \sum_{k^+=1}^{N^+} e^{-E^+_{k^+}L_s}
                   \hat o_{k^+}^\dagger |0_H\rangle t^b \langle 0_H|\hat o_{k^+} 
                       -\sum_{k^-=1}^{N^-} e^{-E^-_{k^-}L_s}
                \hat p_{k^-}^\dagger |0_H\rangle t^b \langle 0_H|\hat p_{k^-} \Bigr\}
 {\cal O}_R[P_R \hat a, \hat a^\dagger P_L] | 0_S\rangle
\nonumber
\end{eqnarray}
where the flavor generator $t^b$ acts on the suppressed flavor indices carried 
by the operators $\hat o_{k^+}^\dagger$, $\hat o_{k^+}$, $\hat p_{k^-}^\dagger$ and $\hat p_{k^-}$.
%
Now we can directly compare the operator appearing in curly
brackets in Eq.~\ref{eq:j5q_me4} (which should reduce to the
operator $2 m_\mathrm{res}\overline{q} t^b\gamma^5 q$ at low
energies) with the second and third terms of the operator in
curly brackets in Eq.~\ref{eq:Gmann2op_2} (which we have shown
does reduce to the operator $m_\mathrm{res}\overline{q} q$ in
this limit).  These two operators are directly related at first
order in $\delta$ by an infinitesmal chiral transformation
where operators in the left-hand factor transform as
\begin{equation}
\hat{o}_{k^+}^\dagger \rightarrow \hat{o}_{k^+}^\dagger (1+i\delta t^b)  \quad\quad
\hat{p}_{k^-}^\dagger \rightarrow \hat{p}_{k^-}^\dagger (1-i\delta t^b)
\end{equation}
while operators in the right-hand factor transform as
\begin{equation}
\hat{o}_{k^+} \rightarrow (1+i\delta t^b) \hat{o}_{k^+} \quad\quad
\hat{p}_{k^-} \rightarrow (1-i\delta t^b) \hat{p}_{k^-}.
\end{equation}
Since this relation between the operators in curly brackets in
Eqs.~\ref{eq:j5q_me4} and \ref{eq:Gmann2op_2} will continue to
hold in the long distance limit, each limiting quantity, 
$2 m_\mathrm{res}\overline{q} t^b\gamma^5 q$ and 
$m_\mathrm{res}\overline{q} q$, must contain the same value
of $m_\mathrm{res}$.  Thus, the residual mass defined in 
Eq.~\ref{eq:mres_axial} agrees with that appearing in the residual 
mass operator deduced from the large $L_s$ limit of the transfer 
matrix for the case $m_f=0$.
This demonstrates the expected consistency with a Symanzik effective 
Lagrangian description of the low-energy effects of mixing between the $s=0$ 
and $s=L_s-1$ walls.

\section{Lattice Actions and Ensembles}
\label{secAction}

Our simulations have been performed with three flavors of dynamical DWF for a class
of improved gauge actions.  Our notation is the same as in
\cite{Blum:2000kn,Aoki:2002vt,Aoki:2004ht}, which we briefly review
here.  The partition function we simulate
is
\begin{equation}
Z = \int [dU] \int \prod_{i=1}^{3} [d\Psi_i d\bar{\Psi}_i]
   \int \prod_{i=1}^{3} [d\Phi^\dagger_{{\rm PV},i} d\Phi_{{\rm PV},i}]
   e^{-S},
\label{eq:Z}
\end{equation}
where the index $i$ runs over the up, down and strange quarks.
The Pauli-Villars fields $\Phi_{{\rm PV},i}$ are needed to cancel
the bulk infinity that would be produced by DWF
as $L_s \to \infty$.  In particular, the total action is
\begin{equation}
S = S_G(U) + S_F(\bar{\Psi}, \Psi, U) +
S_{PV}(\Phi^\dagger, \Phi, U).
\label{eq:total_action}
\end{equation}
The gauge actions we consider are of the form
\begin{equation}
  S_{\rm G}[U] =
   - \frac{\beta}{3} \left[
   (1-8\,c_1) \sum_{x;\mu<\nu} P[U]_{x,\mu\nu} \\
 + c_1 \sum_{x;\mu\neq\nu} R[U]_{x,\mu\nu}\right]
  \label{eq:gauge_action}
\end{equation}
where $P[U]_{x,\mu\nu}$ and $R[U]_{x,\mu\nu}$ represent the real
part of the trace of the path ordered product of link variables
around the $1\times 1$ plaquette and the $1\times 2$ rectangle,
respectively, in the $\mu,\nu$ plane at the point $x$, and $\beta
\equiv 6/g^2$ with $g$ the bare quark-gluon coupling.
We use two common choices for $c_1$: 
\begin{itemize}
\item the Iwasaki action which sets $c_{1} = -0.331$
 \cite{Iwasaki:1985we,Iwasaki:1984cj}; and
\item the DBW2 action which has $c_1 = -1.4069$
\cite{Takaishi:1996xj,deForcrand:1999bi}.
\end{itemize} 
Both are known to 
reduce the residual mass very effectively in the quenched case. 

The fermion action in Eq.~\ref{eq:total_action} can be expressed 
in a form very close to the way it appears in the 3-flavor numerical 
work if it is written as:
\begin{equation}
S_F = - \sum_{i=1}^{3} \bar{\Psi}_i \left[ D_{\rm DWF}^\dagger(M_5, m_i)
  D_{\rm DWF}(M_5, m_i) \right]^{1/2} \Psi_i
\label{eq:fermion_action}
\end{equation}
where $m_i$ is the input bare quark mass for the $i$th light quark 
flavor and the DWF Dirac operator $D_{\rm DWF}$ is defined in
Eq.~\ref{eq:D_dwf}.  We only consider the case where all light quarks have the
same value for the five-dimensional domain-wall height, $M_5$.  The
action for the Pauli-Villars fields is similar, except that the
quark mass $m_i$ is replaced by 1, to yield
\begin{equation}
S_{PV} =
  \sum_{i=1}^{3} \Phi^\dagger_i  \left[ D^\dagger_{\rm DWF}(M_5, 1) D_{\rm DWF}(M_5, 1) \right]^{1/2} \Phi_i
\label{eq:pv_action}. 
\end{equation}
The resulting determinants are suitable for simulation
using a weight  
\begin{equation}
\prod_i
\frac{
 \det^{1/2} \left[ D_{\rm DWF}^\dagger(M_5, m_i) D_{\rm DWF}(M_5, m_i)
   \right] }{
 \det^{1/2} \left[ D_{\rm DWF}^\dagger(M_5, 1) D_{\rm DWF}(M_5, 1)
    \right]
 }
\label{eq:nf3_det}.
\end{equation}
The action may be simulated using either the inexact R algorithm \cite{Gottlieb:1987mq}, 
with associated step size errors,  or 
the exact RHMC algorithm of Clark and Kennedy \cite{Clark:2006fx,Clark:2006wp}.
For the RHMC simulations, each of the square root 
factors in Eq.~\ref{eq:nf3_det} are approximated by a rational
function accurate over some eigenvalue range. 
Thus our simulation uses a 1+1+1 framework that happens to be mass degenerate.
Where our mapping of the Aoki phase finds a clearly non-zero mobility edge it should be 
relevant to our more general programme of 2+1 flavour simulations where a chiral limit
is taken in mass degenerate up/down quark masses at fixed strange mass. 
When written as a partial fraction expansion, each term is individually amenable to the 
usual HMC 
pseudo-fermionic integral (though, of course, a single pseudo-fermion and 
multi-mass inverter is used for all terms in the partial fraction expansion).
We used the R algorithm for the earlier runs in our parameter search
and used the RHMC out of preference, as documented in Table~\ref{tabParams}.
Our implementation of the RHMC used separate fields
as stochastic estimators of the numerator and denominator of Eq.~\ref{eq:nf3_det}.
We simulated all three flavors in the numerator as separate single-flavor species,
and treated the Pauli Villar fields as seperately estimated two- and single-flavour pseudofermions. 
Thus, we 
did not make any explicit use of the mass degeneracy between up and down quarks, (as would have been the case, for example, in two-flavor HMC).  

The ensembles used in this study are a subset of those presented in 
\cite{Antonio:2005yh,Antonio:2005jm,Antonio:2006px}, and have been
used for precision valence measurements with $L_s=8$. We used a
reduced statistic subset to perform a consistent study as we vary $L_s$ in the 
(still expensive) valence analysis. 
The larger statistical sample used for the 
$L_s=8$ data points in \cite{Antonio:2005yh,Antonio:2005jm,Antonio:2006px} lead to lower errors
on results for $m_{\rm res}$ than we present in this work. We include only 
lower precision values in this paper to keep the $L_s = 8 $ 
analysis identical in all systematic respects to the $L_s > 8$ points that are unique to this work.
The ensembles used in this work are listed in Table~\ref{tabParams}.
For the full simulation details, we refer the reader to \cite{Antonio:2005yh,Antonio:2005jm,Antonio:2006px}
where the entire dataset is presented. 

All ensemble and correlation function computations were performed 
using the Columbia Physics System (CPS) on
QCDOC computers in Edinburgh and Brookhaven. All measurements of four-dimensional eigenmodes
were performed using the CHROMA physics system \cite{Edwards:2004sx}, while the 5-dimensional
modes were measured using CPS. The relation between these modes provides a
non-trivial cross-check between code bases.
Both code bases made use of the optimized Bagel QCD library \cite{Bagel}.
Visualization was carried out using the OpenDX software package.

\section{Spectrum of the Hermitian Wilson Dirac Operator}

\label{secEigenmodes}
In this section we study the eigenspectrum and eigenmodes of $H_W$ on our ensembles. 
We determined up to 256 eigenvectors of $H_W$ on 25 configurations per ensemble,
at the negative mass $M_5=1.8$ that was used in our domain wall action for the sea quarks. Additional eigenvalues were
generated on a number of configurations as a function of $M_5$.
Our study therefore includes the dependence of the low-lying spectrum on $M_5$, often called 
spectral flows, in Section~\ref{secFlows}. 
We also study the spectral density of near-zero modes in Section~\ref{secDensity}, 
the microscopic shape distribution of the low-lying modes in Section~\ref{secMobilityEdge},
and the relation between modes of the hermitian 5-dimensional operator 
$H_{\rm DWF} = \gamma_5 R_5 D_{\rm DWF}$ and $H_W$ in Section~\ref{secFourDFiveD}, 
where $R_5$ is the reflection operator in the fifth dimension.  

\subsection{Spectral flows}
\label{secFlows}

Prior to recent work on localization, it was believed that a gap in the 
spectrum was required for the correct behaviour of the various lattice
formulations of chiral symmetry. 
In this context, it was believed that by obtaining spectral flows of the lowest
few modes on a ``typical'' configuration, 
the health of the formulation was indicated by a nicely opened ``eye'' for
some range of $M_5$ containing the Wilson mass used in the simulation.
Recent improved understanding suggests that a non-zero, but low, density of 
localized states exists throughout the region in which we can afford to simulate. 
A gap, as such, does not exist after an ensemble average and, if spectral flows from 
sufficiently many configurations were overlaid, the entire eye would be filled in.

\begin{figure}[hbt]
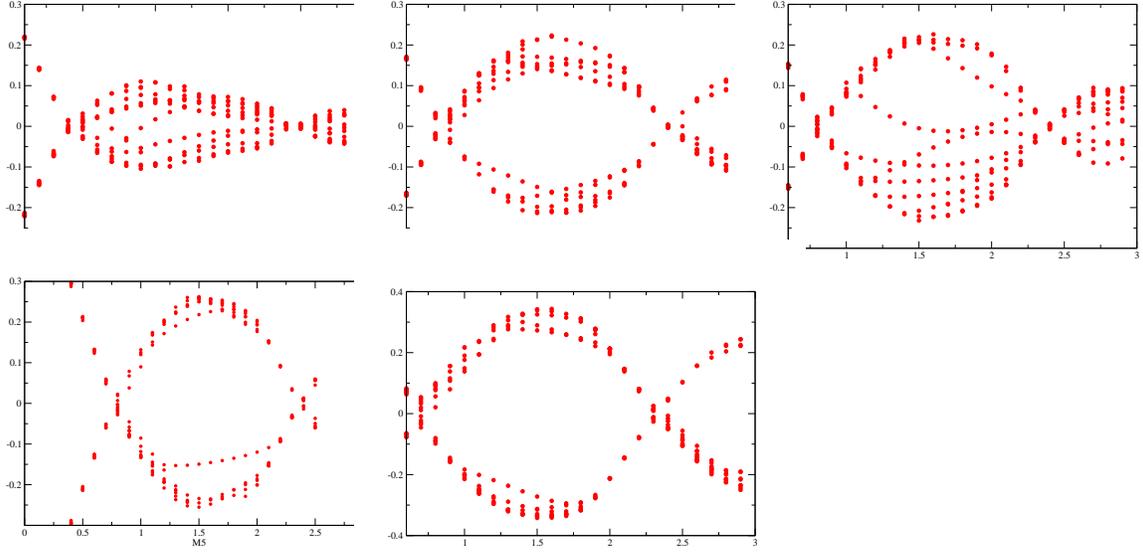

\begin{center}
\begin{tabular}{ccc}
\includegraphics[width=0.3\textwidth]{fig1a_b0_72_sf.eps}  & 
\includegraphics[width=0.3\textwidth]{fig1b_b0_764_sf.eps} &
\includegraphics[width=0.3\textwidth]{fig1c_b0_78_sf.eps} \\
\includegraphics[width=0.3\textwidth]{fig1d_b0_8_sf.eps}   &
\includegraphics[width=0.3\textwidth]{fig1e_b0_88_sf.eps}   
\end{tabular}
\caption{\label{fig:bubbleDBW2} Spectral flow of DBW2. The horizontal axis is
the valence domain wall height, $0\le M_5\le3.0$ and 
the vertical axis is the eigenvalue of 
$H_W$, $-0.3\le\lambda_{H_W}\le0.3$.
From left, $\beta=0.72, , \beta=0.764, \beta=0.78$ (first row) and 
$\beta=0.8,\beta=0.88$ (second row).}
\end{center}
\end{figure}

\begin{figure}
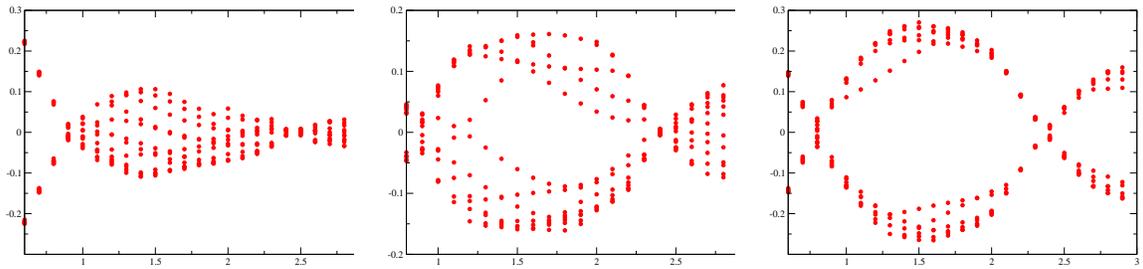

\begin{center}
\begin{tabular}{ccc}
\includegraphics[width=0.3\textwidth]{fig2a_b2_13_sf.eps} &
\includegraphics[width=0.3\textwidth]{fig2b_b2_2_sf.eps}  &
\includegraphics[width=0.3\textwidth]{fig2c_b2_3_sf.eps}
\end{tabular}
\caption{\label{fig:bubbleIWA} Spectral flow of Iwasaki $\beta=2.13, \beta=2.2$, and  $\beta=2.3$ (from left to right).
Axes are the same as fig \ref{fig:bubbleDBW2}.}
\end{center}
\end{figure}

However, on a small sample of configurations an open eye is indicative of a low density 
of low-lying modes and, we argue, a non-zero mobility edge since our results show that
the spectral density grows rapidly across this edge. 
The flows presented here also allow
comparison between our ensembles and those of previous works.
These plots for the lowest modes, Figures~\ref{fig:bubbleDBW2} and \ref{fig:bubbleIWA},
are therefore useful information at the larger values of $\beta$ where the
open eye is indicative of a non-zero lower bound for the mobility edge,
as presented previously~\cite{Antonio:2005wj}. 

Our spectral flow results are certainly indicative of a non-zero mobility edge
for $\beta \ge 0.78$ for the DBW2 gauge action and $\beta \ge 2.2$ for the Iwasaki gauge action.
For smaller values of $\beta$ the spectral flows are inconclusive as no opening is displayed
suggesting a non-zero density of low modes. We shall
return to look at the localization structure in a more sophisticated fashion and classify
these modes in Section~\ref{secMobilityEdge}.

\subsection{Spectral density}

\label{secDensity}

\begin{figure}[hbt]
\includegraphics[width=\textwidth]{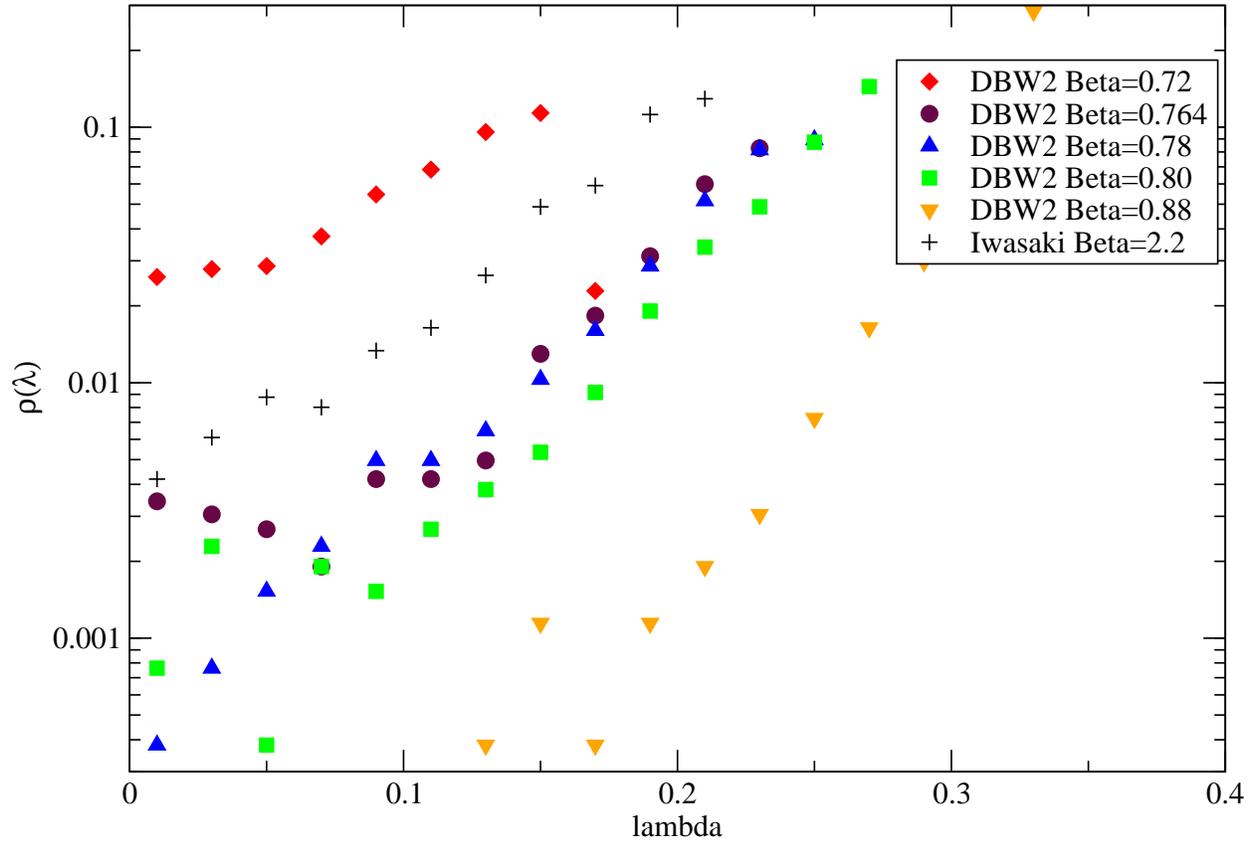}
\caption{\label{figSpectDensity}
Spectral density obtained on a subset of our data with no error analysis, and
fixed bin width $0.02$.
Thus, where the density is low the error could be large and the data
should be considered only as a qualitative indication of the nature of our ensembles.
}
\end{figure}

For each dataset comprising $n_{\rm conf}$ configurations we label the $i$th eigenvalue on the 
$j$th configuration $\lambda_{ij}$.
We use a somewhat simplistic binning approach to estimate the spectral density for a partitioning
of the spectrum. For the $k$th bin we estimate the density as
$$
\rho^k([\lambda^k_{\rm min} + \lambda^k_{\rm max}]/2) = \frac{1}{n_{\rm conf} } \frac{N_k}{\lambda^k_{\rm max} - \lambda^k_{\rm min}},
$$
where $N_k$ is the number of eigenvalues $\lambda_{ij}\in [\lambda^k_{\rm min} + \lambda^k_{\rm max}], j=1,\ldots,n_{\rm conf}$.
Figure~\ref{figSpectDensity} displays our results for the spectral densities for our various datasets, using
a fixed bin width of 0.02.
Some clear, and unsurprising, trends are evident: the spectral density $\rho(0)$ is non-zero, and
falls rapidly as $\beta$ is increased.  The quantity $\rho(0)$ is slightly higher for the Iwasaki 
action at $\beta=2.2$
than for DBW2 at $\beta= 0.78$ which have comparable lattice spacings.  The density
$\rho(\lambda)$ rises rapidly, and perhaps exponentially, in $\lambda$ as we move away from zero.

\subsection{Microscopic study of the mobility edge}
\label{secMobilityEdge}

\begin{figure}[hbt]
\includegraphics[width=\textwidth]{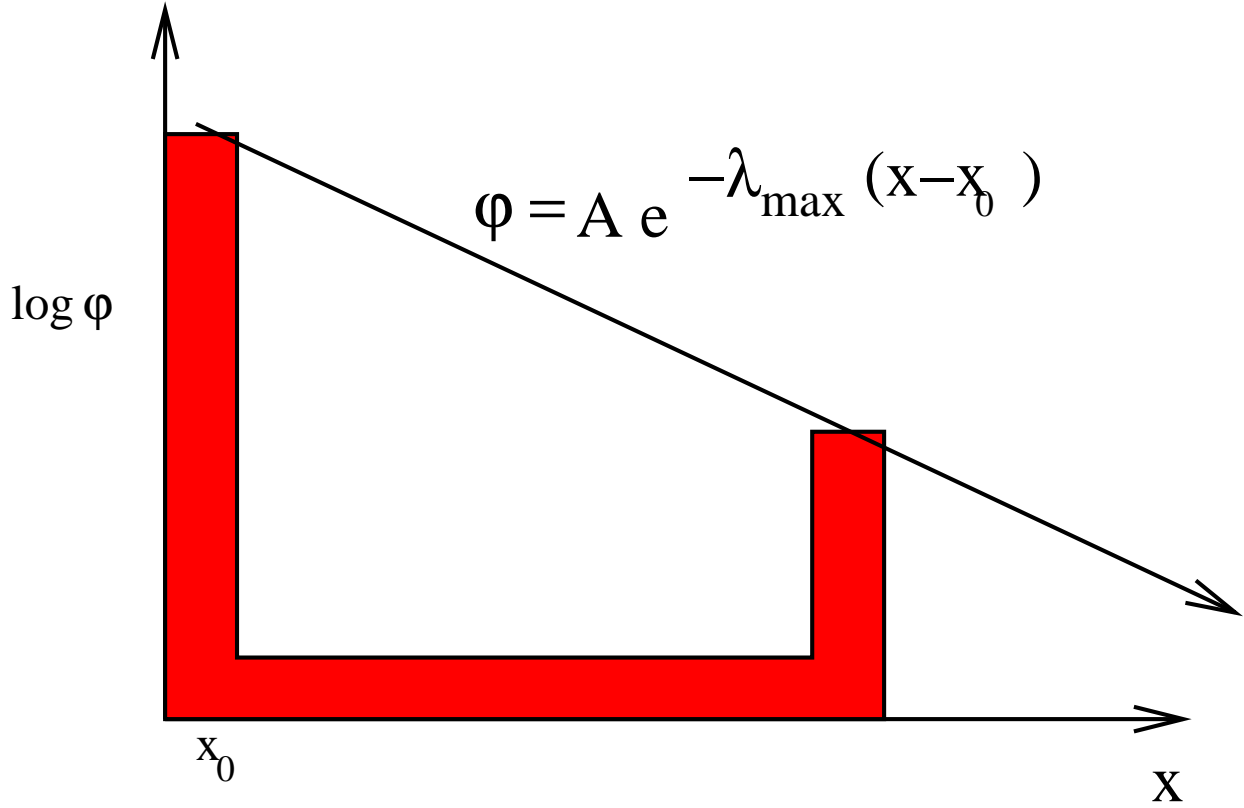}
\caption{\label{figLocalisationMeasures}
Various measures can be used to define the degree of localization.
Inverse participation ratio has been a popular order parameter for delocalization
transitions and, while fine for that purpose, counts occupied sites without paying attention
to mode shape. This one dimensional cartoon highlights the
advantages of the more robust measure employed in this paper. Given the model
of exponential localization, our tactic is to find a localization exponent that forms a tight
bound for the eigenvector. This measure will show the influence of satellite peaks in a robust 
fashion by eliminating all forms of dilution by volume average. 
Other useful measures that take into account the shape would include appropriately 
weighted moments of the density function.
} 
\end{figure}

We wish to study the size distribution of our eigenvectors, $\psi(x)$, as a function of the eigenvalue
in order to test the mobility edge conjecture. This approach is novel and was first presented 
in \cite{Antonio:2005wm} for dynamical DWF, with a similar approach taken in \cite{Yamada:2006fr} for dynamical overlap.
A standard measure is the inverse participation ratio, 
$$P^{-1} = \sum\limits_x |\psi^\dagger(x)\psi(x)|^2 .$$
A normalized extended state will produce $P\simeq V$, while a completely localized
state will produce $P\simeq 1$. Thus the (inverse) participation ratio essentially counts
the number of occupied sites and, in the infinite volume limit,
is a useful order parameter for delocalization transitions.
However, this metric is not sensitive to the shape of a state and so  
makes no distinction between occupying two adjacent sites, and occupying two well
separated sites, such as the illustrative non-local example in 
Figure \ref{figLocalisationMeasures}. 

We want to understand how to reliably distinguish exponentially compact states from those with \emph{any}
degree of long-range support. In the infinite volume limit the use of the inverse participation ratio
only identifies long-range correlations arising from structures whose four dimensional
volume diverges and we seek a more robust measure.

While a better measure could be a moment of the eigenvector density, such as
$$\sum\limits_x  \psi^\dagger(x)\psi(x) |x|  ,$$
we have found empirically, by visualizing data, that near the delocalization threshold the
dominant long range correlation arises from multi-peaked eigenvectors, which fall exponentially
around multiple centers. We define a more robust measure of exponential localization length,
which generalizes the one dimensional case of 
Figure~\ref{figLocalisationMeasures} to higher dimensions as follows

\begin{itemize}
\item The mode density is defined as $\phi(x) = \psi^\dagger(x) \psi (x)$. 
\item We identify a center $x_0$ such that $\phi(x_0) \ge \phi(x) \forall x \ne x_0$.
\item For every site, we define a radius $r(x) = | x-x_0 |$ using the periodic mirror image nearest to $x_0$.
\item We define an effective localization exponent $L_{\rm eff}(x) = \frac{2 r(x)} {\log \phi(x_0) - \log \phi(x)}$. 
\item Finally, we define a robust localization length $L_{\rm max} = \max\limits_{r(x)\ge5} L_{\rm eff} (x)$.
\end{itemize}

\begin{figure}[hbt]
\includegraphics[width=\textwidth]{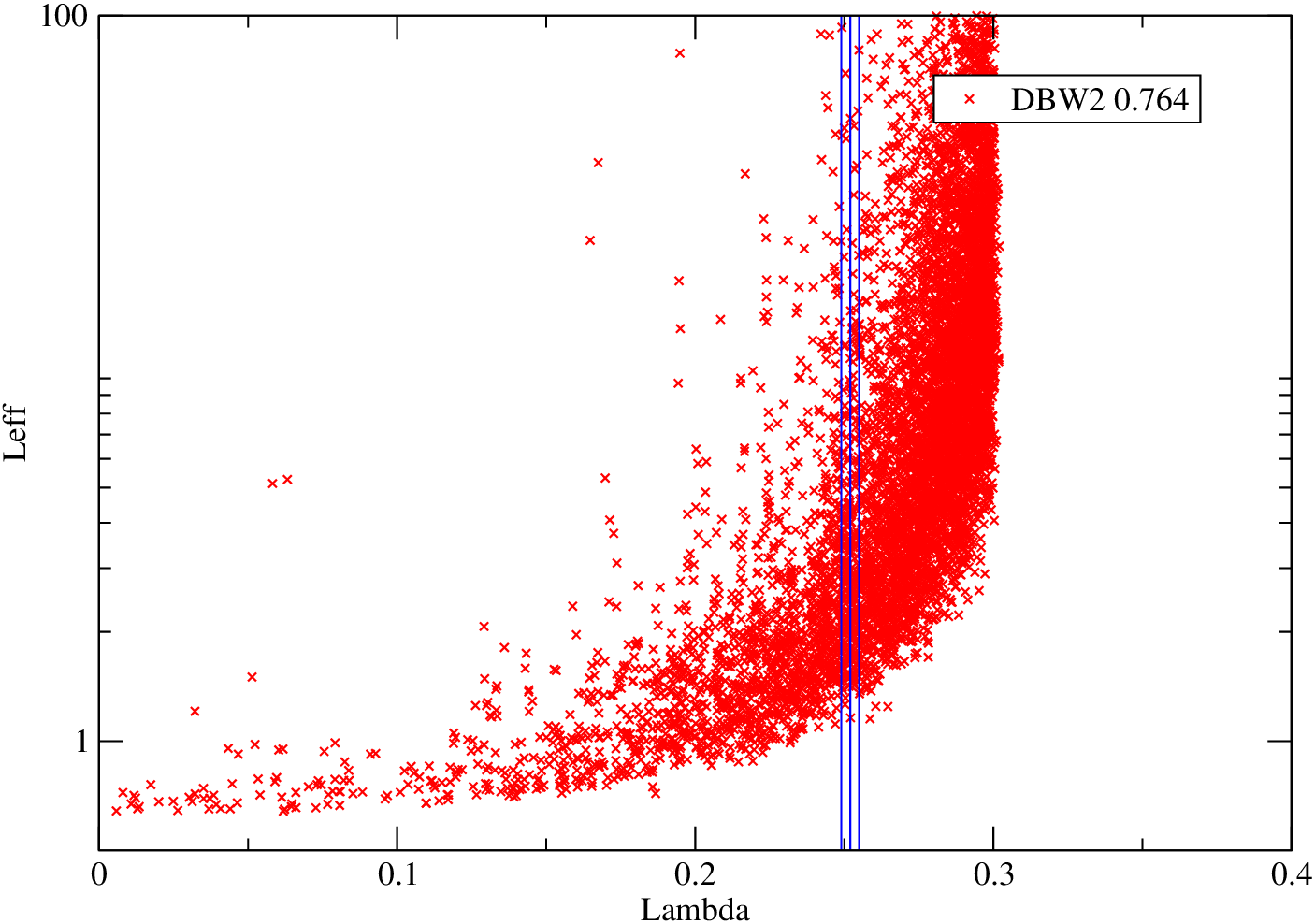}
\caption{\label{MicroscopicMobilityDBW2}
Microscopic view of the mobility edge based on a scatter plot of maximal localization lengths
of individual low eigenmodes of $H_W$. We overlay the $H_T$
mobility edge $\lambda_c$ from fits to the model for $m_{\rm res}$, Eq~\ref{eq:mresLs}
and find good agreement with our diverging localisation length 
for the DBW2 gauge action at $\beta=0.764$.
}
\end{figure}

\begin{figure}[hbt]
\includegraphics[width=\textwidth]{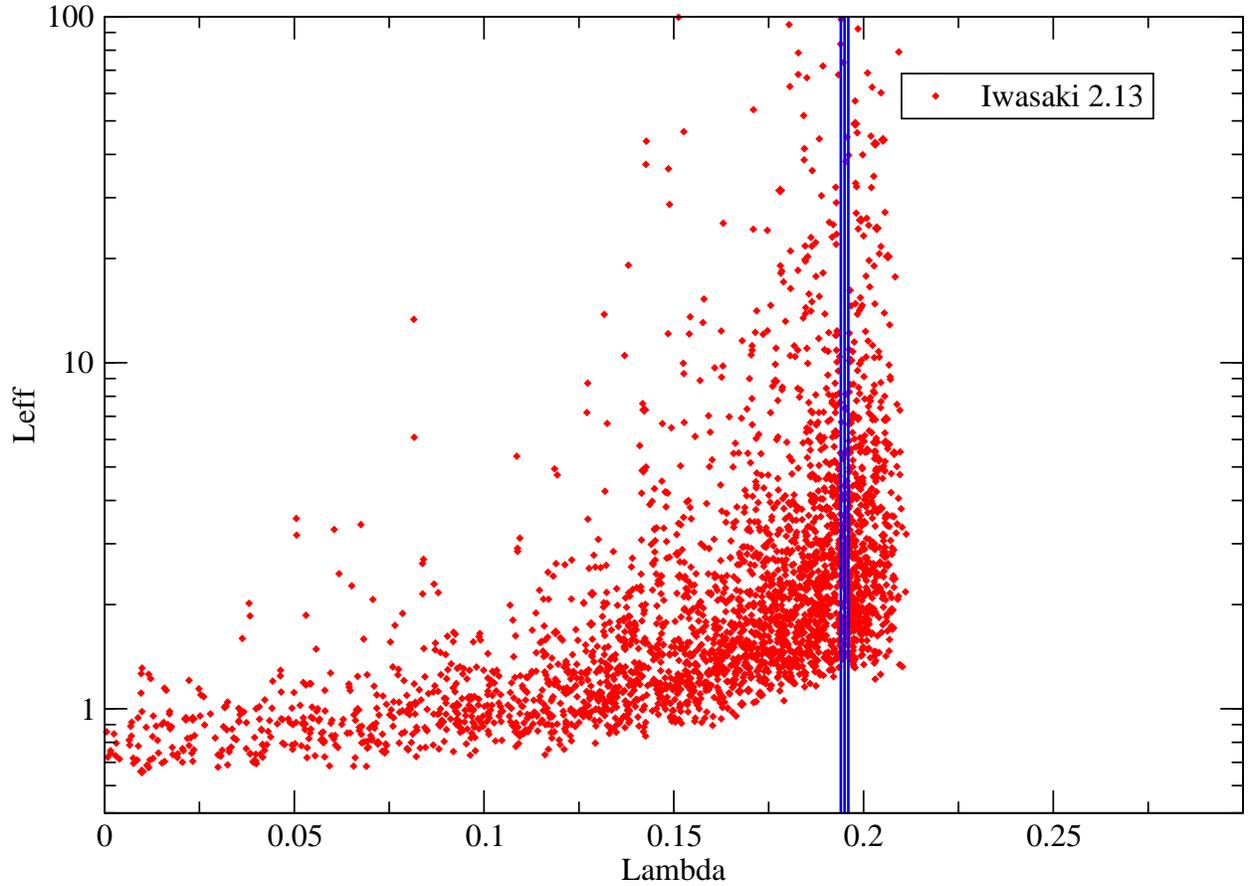}
\caption{\label{MicroscopicMobilityIwasaki}
Microscopic view of the mobility edge based on a scatter plot of maximal localization lengths
of individual low eigenmodes of $H_W$. We overlay the $H_T$
mobility edge $\lambda_c$ from fits to the model for $m_{\rm res}$, Eq~\ref{eq:mresLs}
and find good agreement with our diverging localisation length 
for the Iwasaki gauge action at $\beta=2.13$.
}
\end{figure}

As the eigenvalue approaches the mobility edge from
below, we observed two clear processes by which delocalization takes place. Firstly,
the mean rate of fall off from the center decreases creating much bigger
states. Secondly, states become increasingly multi-centered, with exponential fall
off between a number of satellite peaks. The above definition of 
localization length is designed to reflect both the process of single-peak broadening,
and the development of multiple peaks. The inverse participation ratio would
be a less robust measure of locality in that it would not differentiate two well-separated $\delta$ 
functions from those on two neighbouring sites, or indeed would not identify a lower-dimensional
extended sheet in the infinite volume.
We display scatter plots of this definition of localization length 
for the 256 lowest eigenvectors of $H_W$ on 25 configurations per ensemble
in Figure~\ref{MicroscopicMobilityDBW2} and \ref{MicroscopicMobilityIwasaki}
for the DBW2 and Iwasaki gauge actions at $\beta = 2.13$ and $\beta = 0.764$ respectively.

\subsection{Relation between eigenmodes of the five and four-dimensional operators}
\label{secFourDFiveD}

We computed the lowest 10 modes of the 5-dimensional hermitian
domain wall operator $H_{\rm DWF} = \gamma_5 R_5 D_{\rm DWF}$, using the
Columbia Physics System, on a single configuration from a $16^3\times 32\times 8$
DBW2 three flavor ensemble at $\beta=0.764$. 
On the same configuration, we used the lowest 256 eigenmodes $\psi_j(x)$
of $H_W$, computed using the CHROMA software package,
to express the 5-dimensional modes $\Psi_i(x,s)$ as a sum over 4-dimensional modes using
\begin{equation}
\Psi_i(x,s) = N_s \sum_j \alpha_{ij}(s) \psi_j(x),
\label{eq:5_4_expansion}
\end{equation}
where $$N_s^2 =\sum_x \Psi_i^\dagger(x,s) \Psi_i(x,s) $$ normalizes 
each $s$-slice of the 5-dimensional eigenvector to unity.
This means that if the basis of 4-dimensional eigenvectors were complete
\begin{equation}
\sum_j \alpha_{ij}(s)^2 = 1.
\end{equation}
However, since we are only using the lowest 256 eigenmodes, we expect that 
Eq.~\ref{eq:5_4_expansion} will, at best, be true only deep in the bulk for 
large $L_s$ where low modes of the transfer matrix dominate, and only up 
to differences between $H_T$ and $H_W$.

We therefore computed 5-dimensional modes for a valence $L_s = 16$
(N.B. not the unitary $L_s=8$ case) finding
that, while the basis is very much incomplete on the wall, the description
it gives in the 5-dimensional bulk is as much as 80\% complete by the mid-point, despite the
mismatch between $H_T$ and $H_W$. (This also serves as a useful cross check between the eigensolvers
in the two independent code bases.)
 
\begin{figure}
\includegraphics[width=1.0\textwidth]{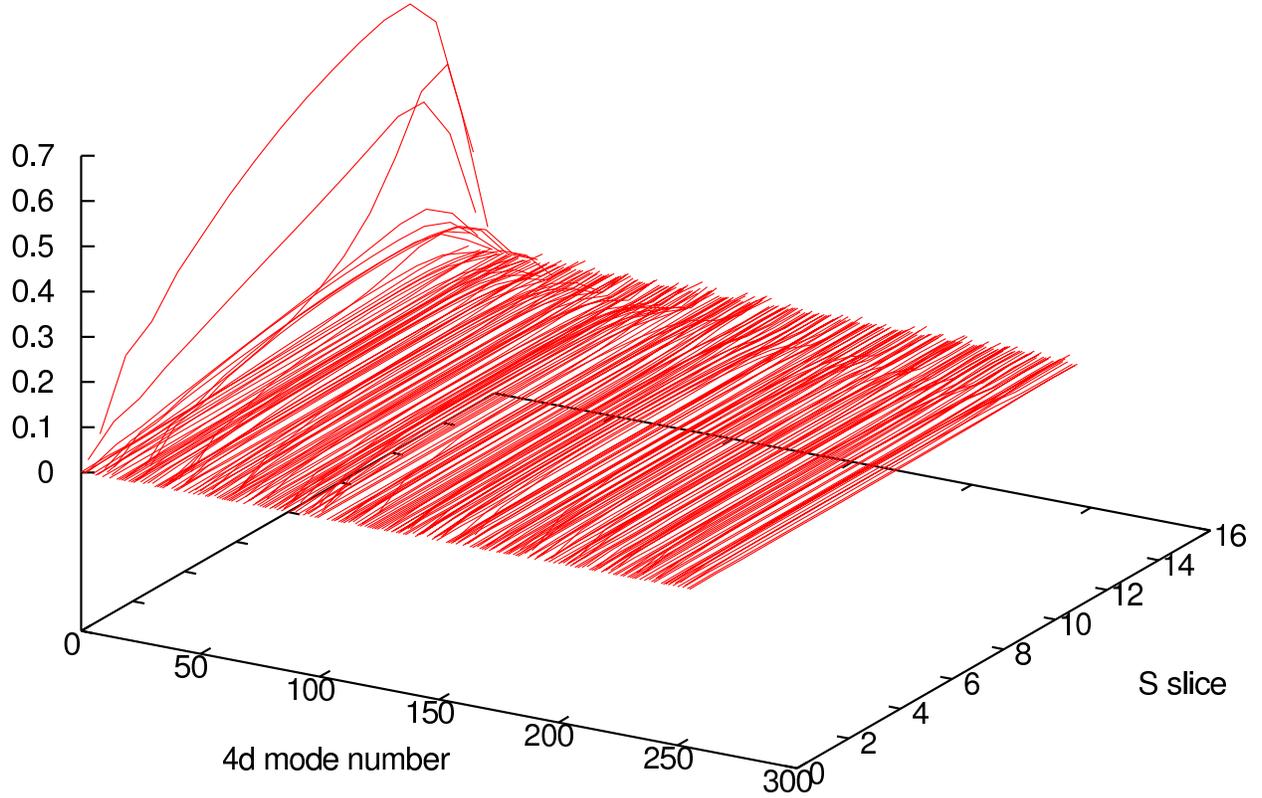}
\caption{\label{DotProduct} Normalized overlap of different $s$-slices of a chiral 5-dimensional eigenmode
of $H_{DWF}$ with the first 256 4-dimensional eigenmodes of $H_W$. We take the valence operator with $L_s=16$,
and demonstrate numerically that only a few low-lying 4-dimensional eigenmodes dominate
the large $L_s$ contributions to chirality mixing operators such as $m_{\rm res}$ by plotting the 
decomposition of the lowest 5-dimensional mode $|\alpha_{0j}(s)|^2$ 
as a function of fifth coordinate $s$, and 4-dimensional mode number $j$.}
\end{figure}

The coefficients $|\alpha_{ij}|^2$ for a typical 
5-dimensional chiral mode are displayed in Figure~\ref{DotProduct}
and it can be seen that only 3 low-lying modes of $H_W$ dominate the $s$-dependence
of eigenmodes of $H_{\rm DWF}$.
Thus, we clearly demonstrate from numerical data that a 
few localized low-lying 4-dimensional modes of $H_W$
almost completely describe the coupling between walls for each 5-dimensional mode of the hermitian DWF operator for
large $L_s$. As discussed, these modes occur with a low 
density - perhaps a ``dilute gas'' is an
appropriate picture - but, for sufficiently large $L_s$, they
dominate the contributions to chiral symmetry
breaking due to exponential suppression of the extended modes 
above the mobility edge, see Eq.~\ref{eq:spec_density_2}.

\begin{figure}[hbt]
\includegraphics[width=\textwidth]{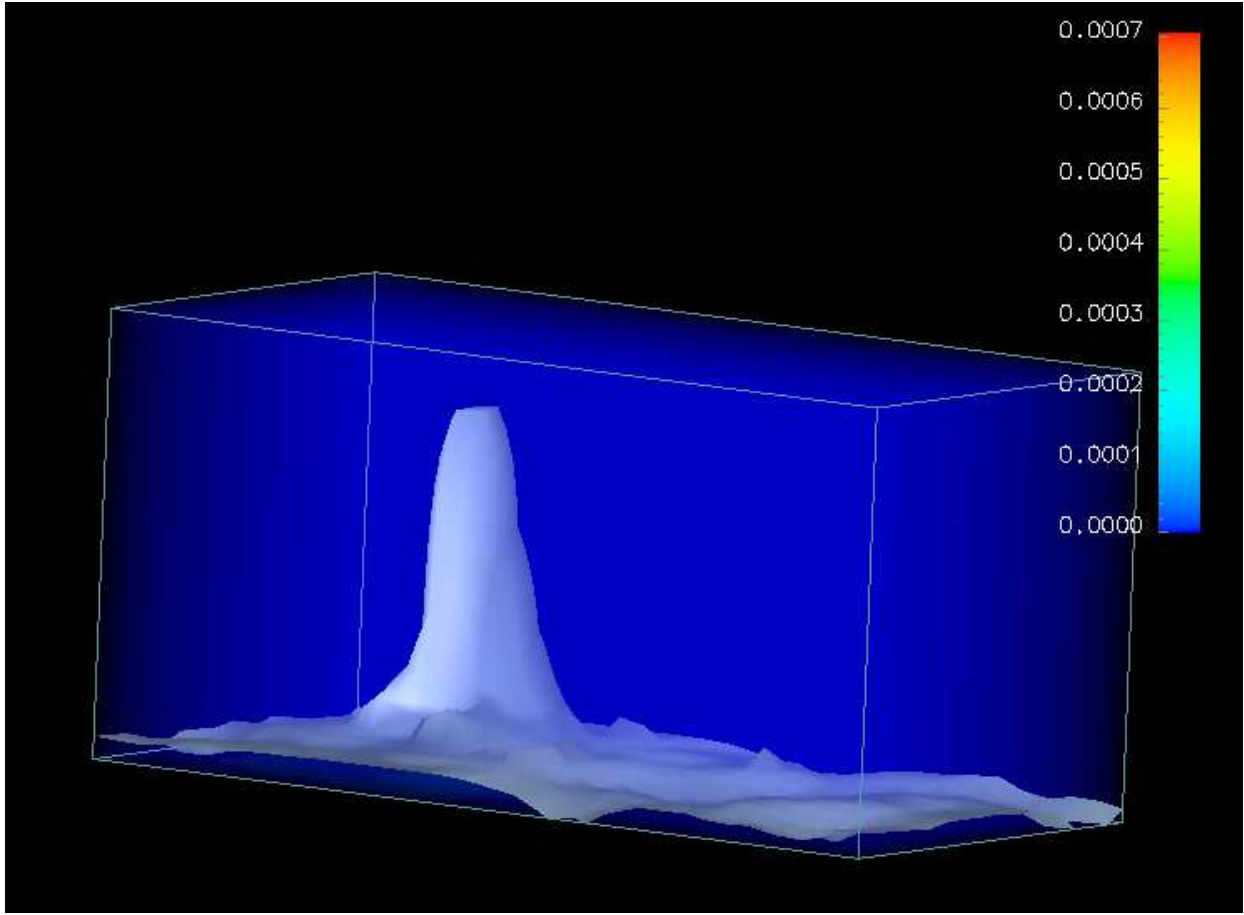}
\caption{\label{5dmodeUnbound}
We display an iso-surface through the eigenvector density of a 
typical near-zero chiral mode of the 5-dimensional operator
$H_{\rm DWF}$, bound to a single wall.
Here $s$ is the vertical direction, $t$ the left-right axis, and $z$ runs into the page.
We show $(x,y) = (3,3)$ showing a localized incursion 
of the eigenmode into the fifth dimension associated with a low mode of $H_W$.
This mechanism is responsible for one form of the contributions to chiral symmetry breaking
in DWF.
}
\end{figure}

\begin{figure}
\includegraphics[width=\textwidth]{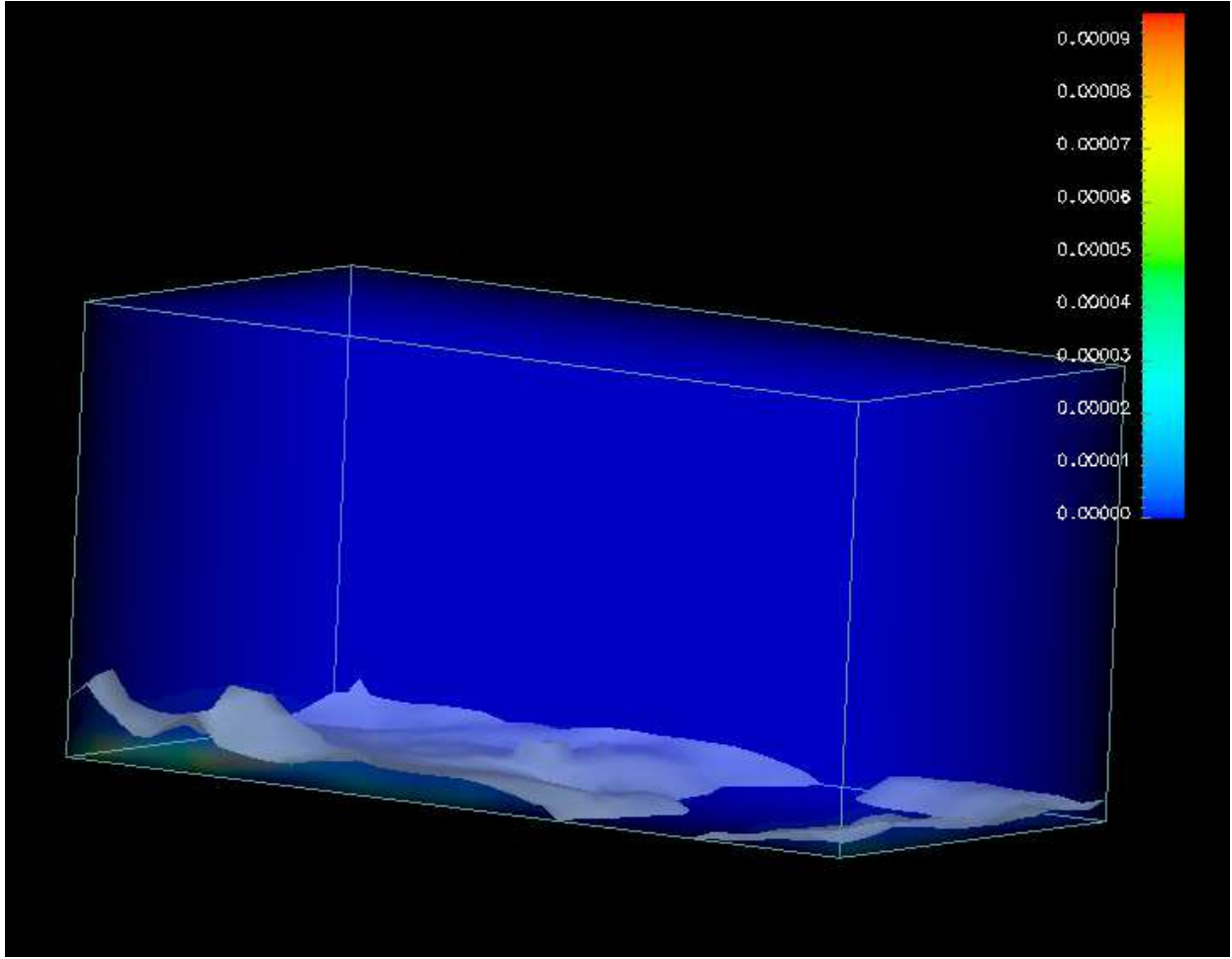}
\caption{\label{5dmodeBound}
We display an iso-surface through the eigenvector density of the same chiral mode displayed in
Fig~\ref{5dmodeUnbound}.
Here $s$ is the vertical direction, $t$ the left-right axis, and $z$ runs into the page.
We show $(x,y)=(11,13)$ showing a uniformly well stuck region of this mode. 
Compared with the previous figure, this degree of binding is the more common
case for this typical eigenvector, and one has to search spatially to 
find any localised incursion.
}
\end{figure}

The localized low modes of $H_W$ thus play a role in providing
conduits into the bulk for chiral symmetry breaking,
causing corresponding localized spikes in the correlation function for $m_{\rm res}$.
This is corroborated by the iso-surfaces (produced with the OpenDX package) of 
constant $H_{\rm DWF}$-eigenvector density (i.e. surfaces defined by
$\Psi^\dagger(x,s)\Psi(x,s) = {\rm constant}$) obtained in our
simulation. Examples of this are shown in Figures~\ref{5dmodeUnbound} and \ref{5dmodeBound},
which represent lower dimensional views, at different spatial locations, of the same eigenvector.
A movie, raster scanning $x$ and $y$ as a unified ``movie-time'', can be 
obtained from 
\footnote{\href{http://www.ph.ed.ac.uk/\~paboyle/QCD/5dmode.mpg}
{http://www.ph.ed.ac.uk/$\sim$paboyle/QCD/5dmode.mpg}}.

Since the low modes of $H_W$ are scarce and very localized,
it is to be expected that if we are in a region
of parameter space where their contribution is the dominant contribution to chiral symmetry
breaking, quantities such as $m_{\rm res}$ will display 
larger non-Gaussian tails (or ``lumpiness'')
than if the exponentially falling contribution from extended states is dominant. 
Thus, one should not be surprised by the need for very high statistics
to achieve a convincing estimate for $m_{\rm res}$ in this 
region of parameter space.

\section{Results for the residual mass}
\label{secMresResults}

\begin{figure}
\includegraphics[width=\textwidth]{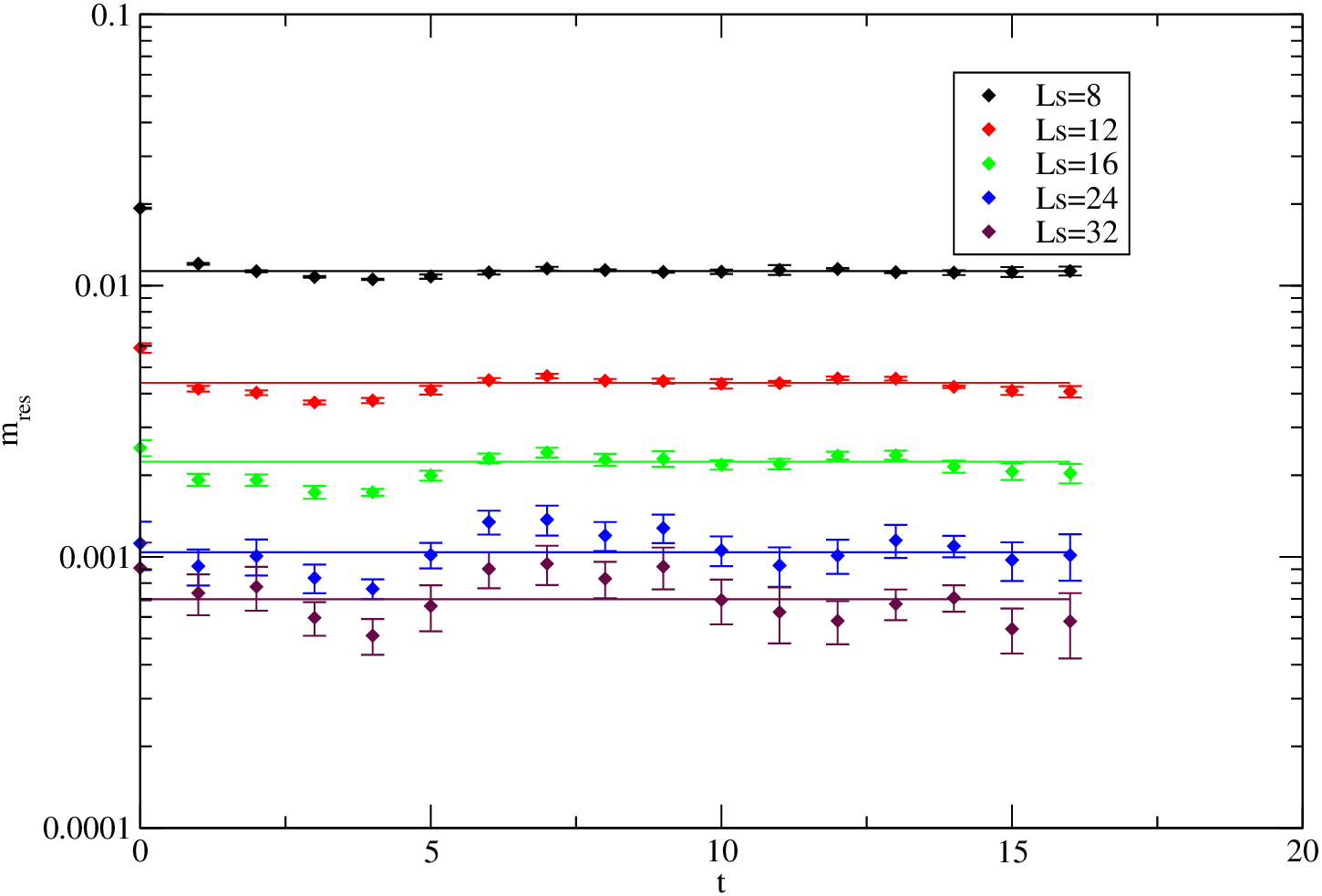}
\caption{
\label{FitMres0}
Residual mass fits for valence quarks with $m=0.04$ and 
$L_s = 8,12,16,24,32$ on the DBW2 $\beta=0.72$ ensemble
with sea quark masses $m_{ud} = m_s = 0.04$, 
and for the sea quark $L_s=8$.
}
\end{figure}

\begin{figure}
\includegraphics[width=\textwidth]{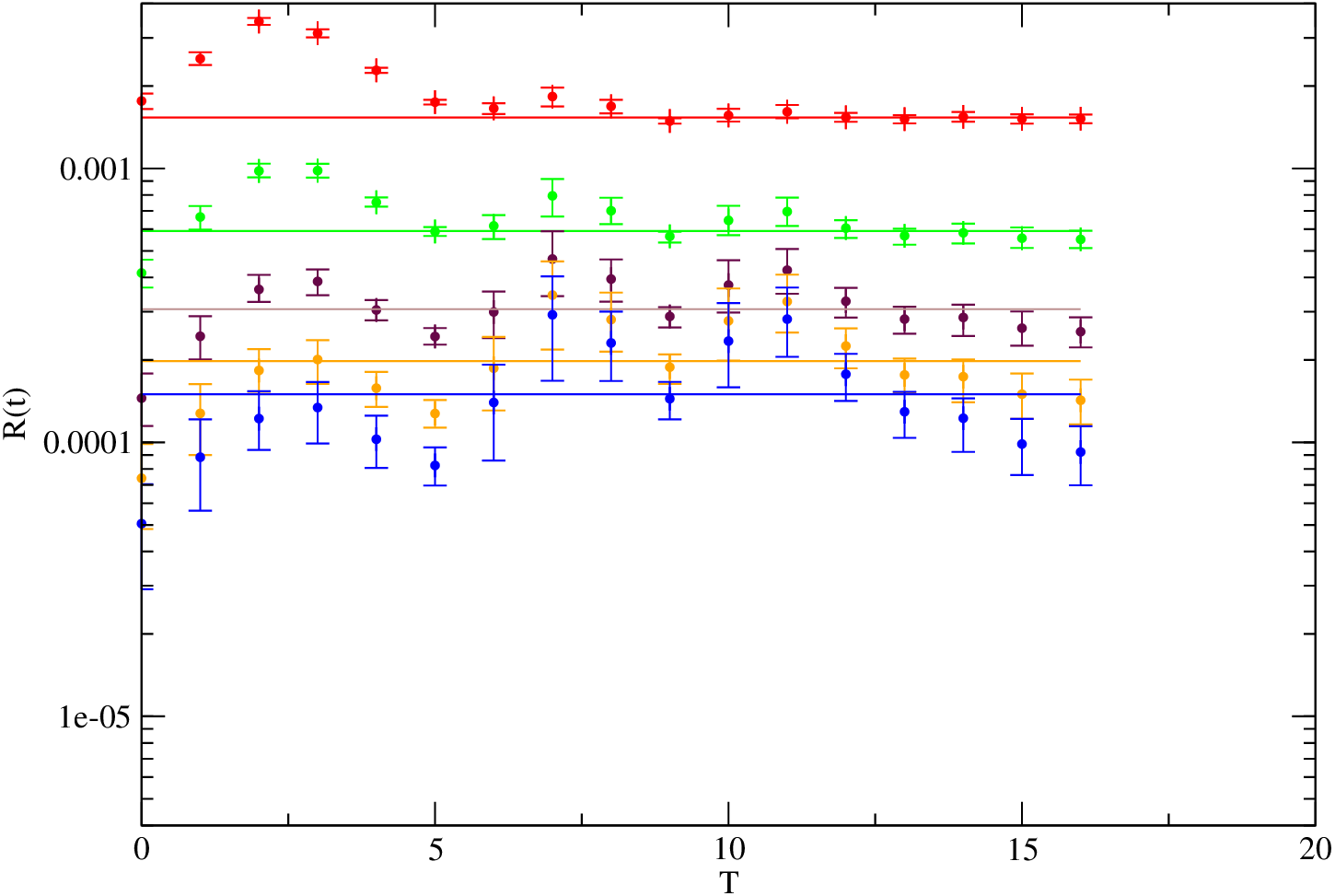}
\caption{
\label{FitMres1}
Residual mass fits for valence quarks with 
$m=0.04$ and $L_s = 8,12,16,24,32$ on the DBW2 $\beta=0.764$ ensemble
with sea quark masses $m_{ud} = m_s = 0.04$ and for the sea quark $L_s=8$.
}
\end{figure}

\begin{figure}
\includegraphics[width=\textwidth]{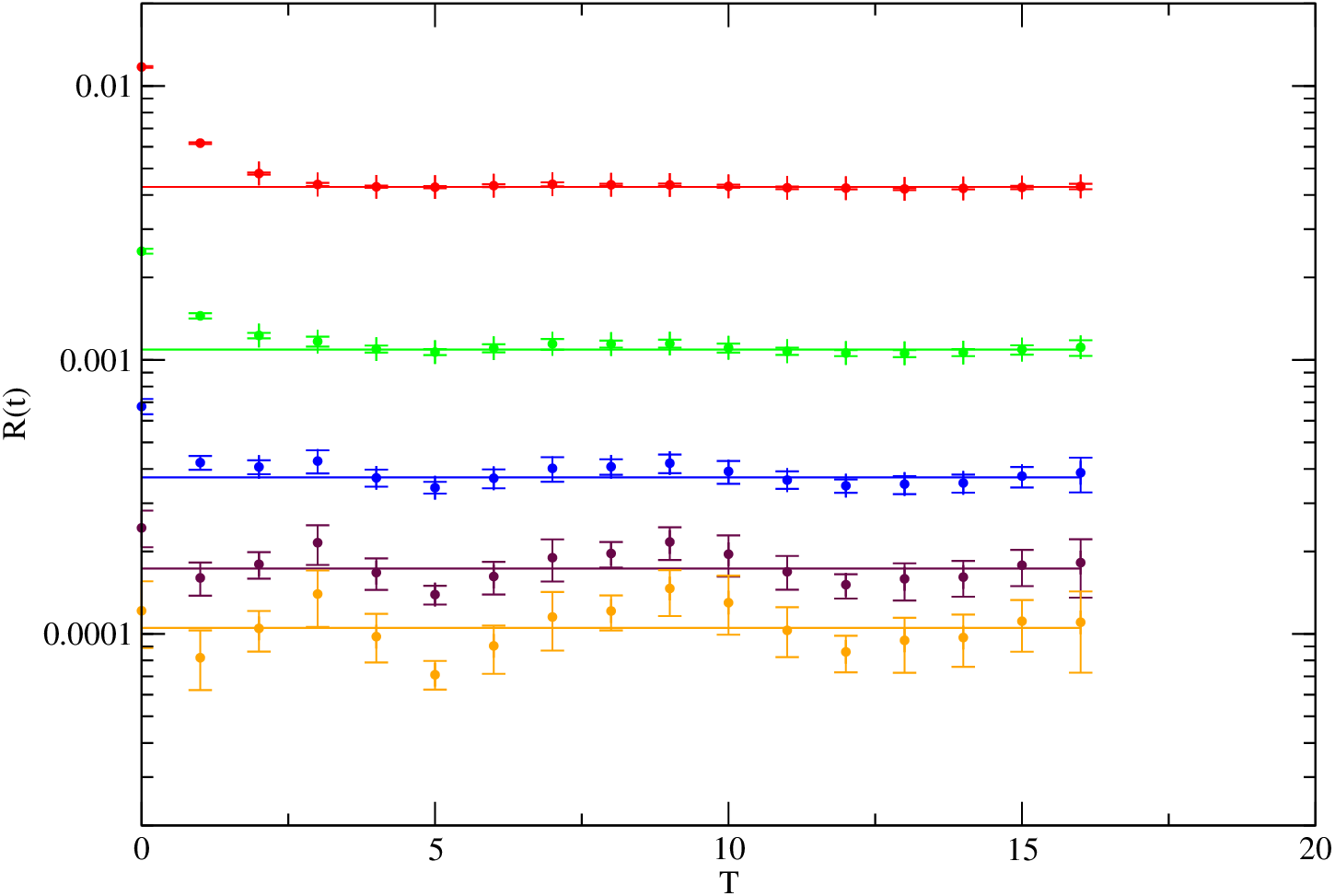}
\caption{
\label{FitMres2}
Residual mass fits for valence quarks with $m=0.04$ and $L_s = 8,12,16,24,32$ on the DBW2 $\beta=0.78$ ensemble
with sea quark masses $m_{ud} = m_s = 0.04$ and for the sea quark $L_s=8$.
}
\end{figure}

\begin{figure}
\includegraphics[width=\textwidth]{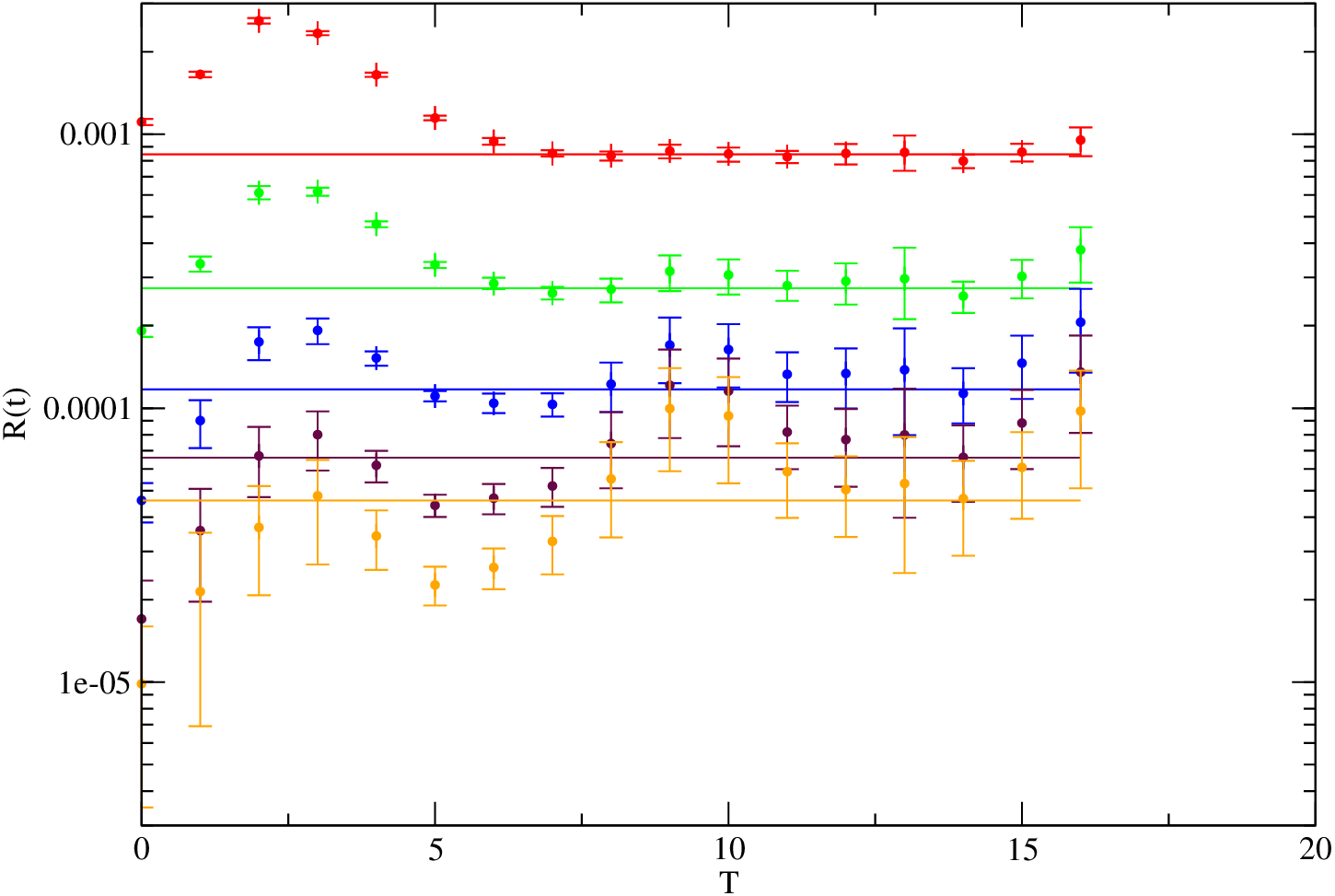}
\caption{
\label{FitMres3}
Residual mass fits for valence quarks with $m=0.04$ and $L_s = 8,12,16,24,32$ on the DBW2 $\beta=0.80$ ensemble
with sea quark masses $m_{ud} = m_s = 0.04$ and for the sea quark $L_s=8$.
}
\end{figure}

\begin{figure}
\includegraphics[width=\textwidth]{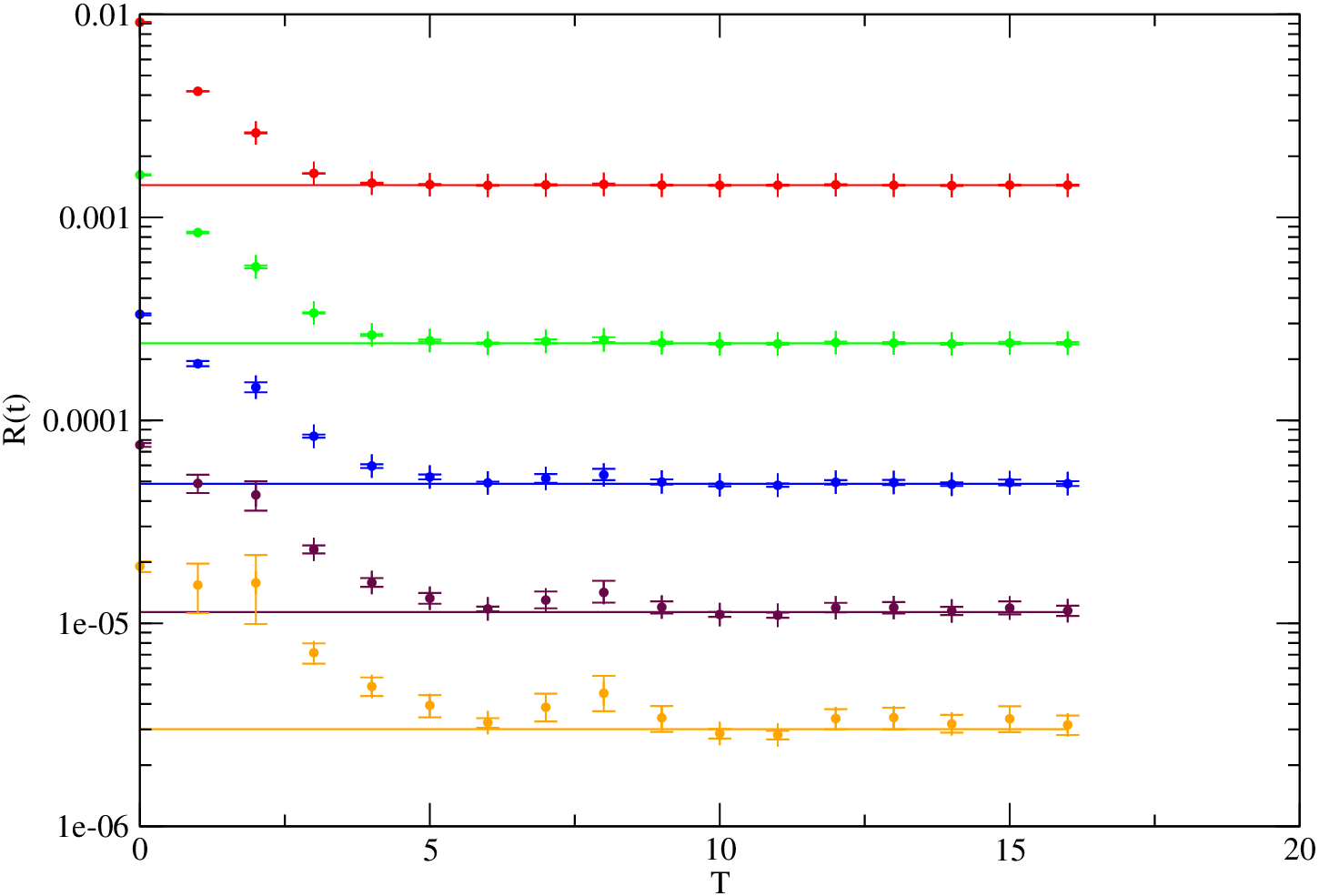}
\caption{
\label{FitMres4}
Residual mass fits for valence quarks with $m=0.04$ and $L_s = 8,12,16,24,32$ on the DBW2 $\beta=0.88$ ensemble
with sea quark masses $m_{ud} = m_s = 0.04$ and for the sea quark $L_s=8$.
}
\end{figure}

\begin{figure}
\includegraphics[width=\textwidth]{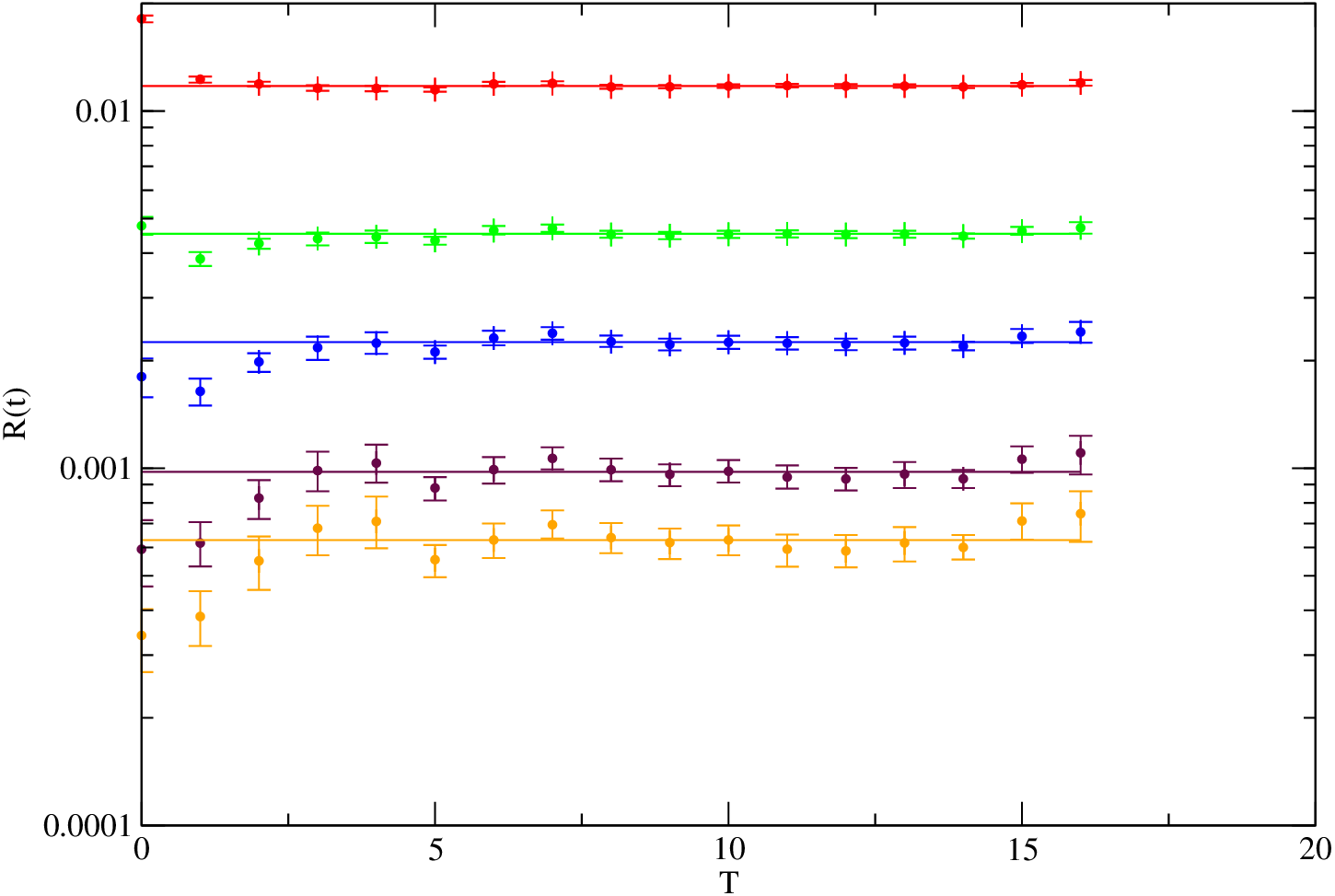}
\caption{
\label{FitMres5}
Residual mass fits for valence quarks with $m=0.04$ and $L_s = 8,12,16,24,32$ on the Iwasaki $\beta=2.13$ ensemble
with sea quark masses $m_{ud} = m_s = 0.04$ and for the sea quark $L_s=8$.
}
\end{figure}

\begin{figure}
\includegraphics[width=\textwidth]{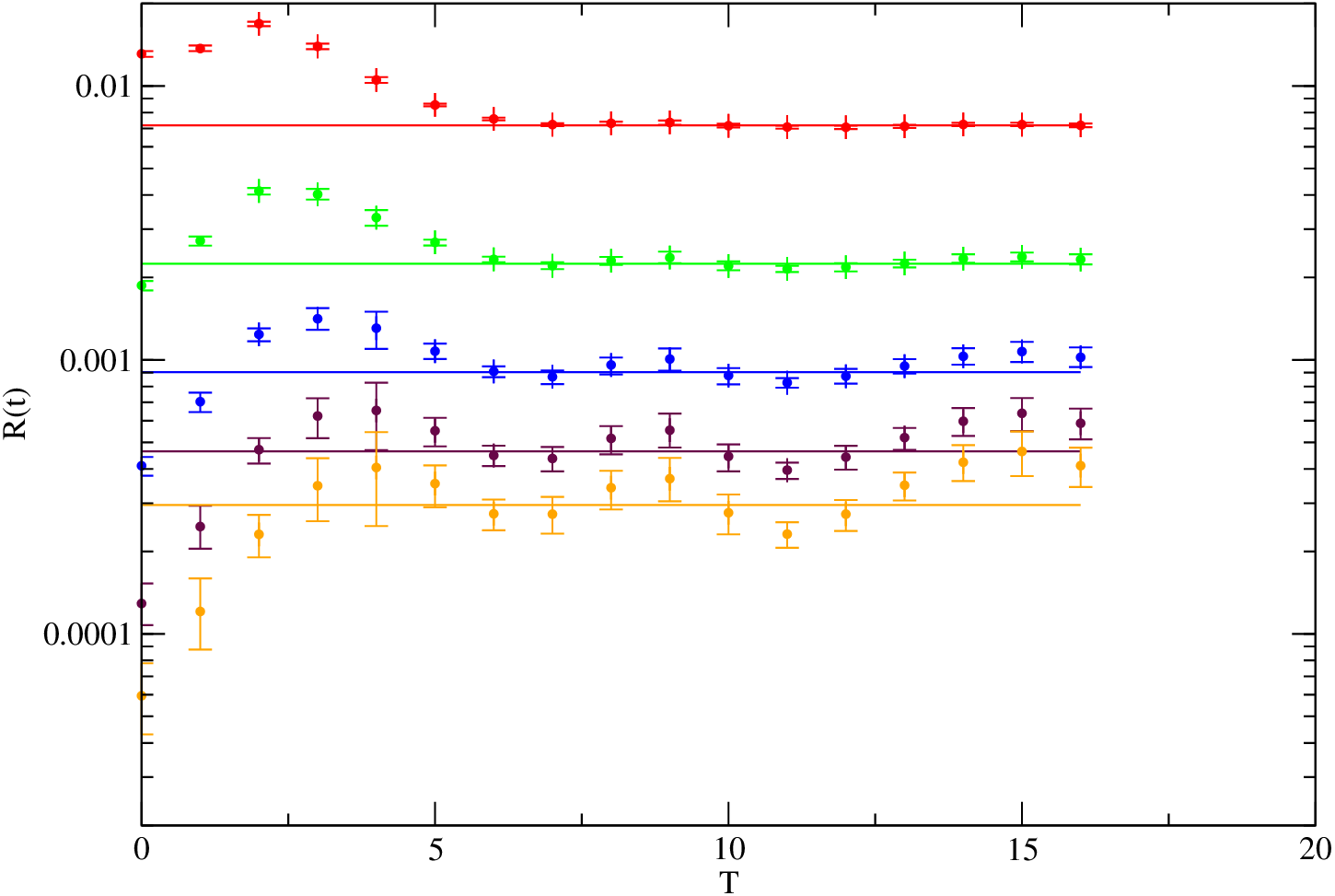}
\caption{
\label{FitMres6}
Residual mass fits for valence quarks with $m=0.04$ and $L_s = 8,12,16,24,32$ on the Iwasaki $\beta=2.2$ ensemble
with sea quark masses $m_{ud} = m_s = 0.04$ and for the sea quark $L_s=8$.
}
\end{figure}

\begin{figure}
\includegraphics[width=\textwidth]{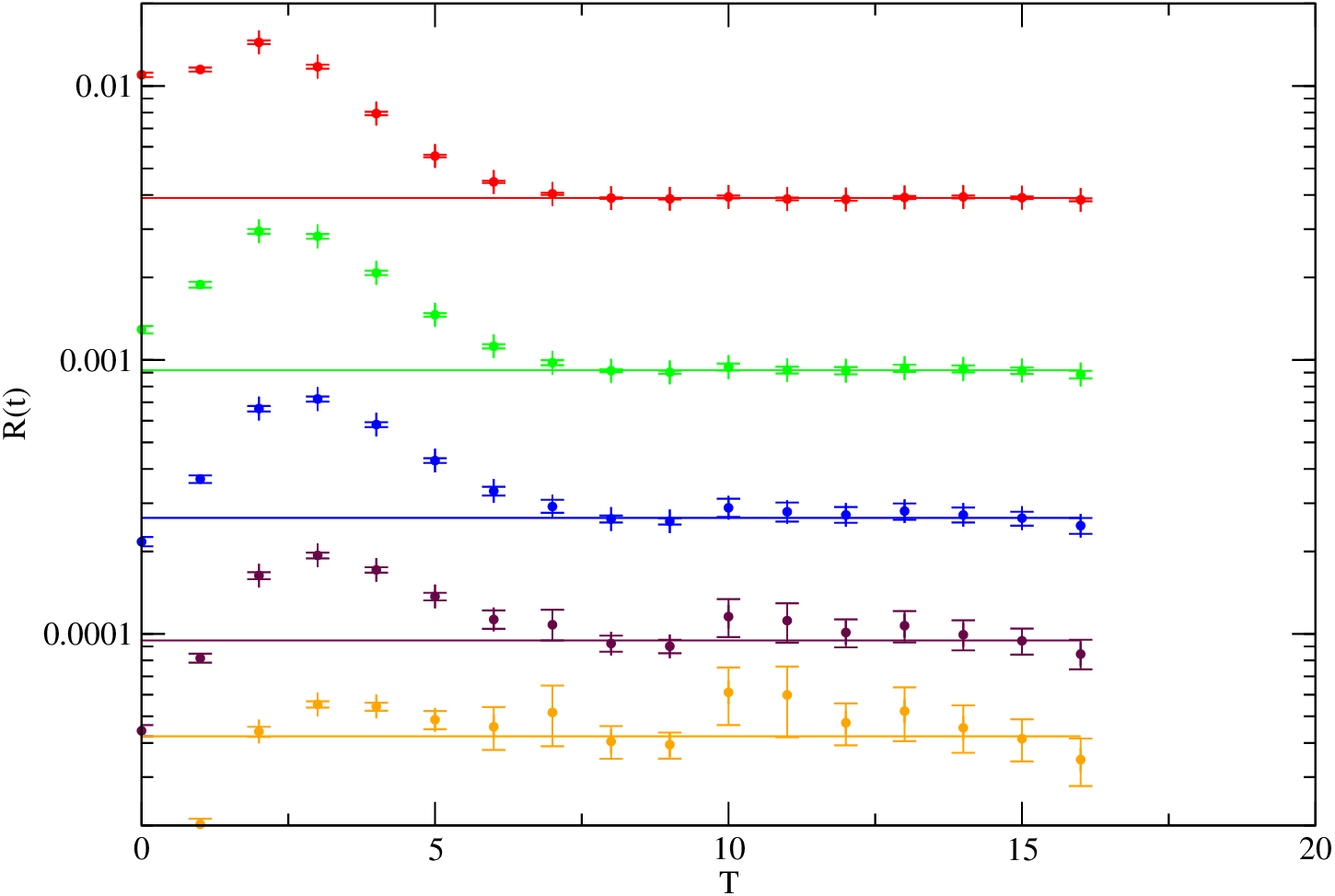}
\caption{
\label{FitMres7}
\label{FitMresN}
Residual mass fits for valence quarks with $m=0.04$ and $L_s = 8,12,16,24,32$ on the Iwasaki $\beta=2.3$ ensemble
with sea quark masses $m_{ud} = m_s = 0.04$ and for the sea quark $L_s=8$.
}
\end{figure}

As discussed in Section~\ref{secMres}, we calculate the residual mass from a
ratio of correlation functions containing the axial Ward identity defect, 
$J^b_{5q}$, defined in Eq.~\ref{eq:j5q_def}:
\begin{equation}
 R(t) = \frac{\sum_{\vec{y}}
 \langle J^b_{5q}(\vec{y},t) P^b(0) \rangle}
    {\sum_{\vec{y}} \langle J^b_5(\vec{y},t) P^b(0) \rangle}
\label{eq:mres_meas}
\end{equation}
where $b$ is a fixed flavor index, and $J^b_5(\vec{y},t)$ is the local 4-dimensional 
pseudoscalar density defined on the $s=0$ and $s=L_s-1$ walls.   The operator 
$P^b(0)$ is either $J^b_5(\vec{y},t)$ or a spatially-smeared variant defined at 
$t=0$.  The residual mass is determined from a constant fit to $R(t)$ 
over a range of times, $t$, that are sufficiently large that lattice artifacts 
should be absent.  Recall that the ratio $R(t)$ should equal the residual 
mass provided that the numerator and denominator of Eq.~\ref{eq:mres_meas} 
are dominated by physical states.  There is not a requirement that only a 
specific lowest state contributes.  For the calculations reported here this 
average is performed over the range $7 \le t \le 16$.  Figure~\ref{FitMres0} 
through Figure~\ref{FitMresN} contain plots of the fits of $R(t)$ to a constant.
We use a combination of point, Coulomb gauge-fixed Wall 
and Coulomb gauge-fixed exponential sources on the various datasets listed 
in Table~\ref{TabMresData}, and display the fitted values for $m_{\rm res}$ 
in Table~\ref{TabMresDBW2} and Table~\ref{TabMresIwasaki} for the DBW2 and 
Iwasaki gauge actions respectively.

In order to conserve computer resources, we studied the dependence of $m_{\rm res}$
on $L_s$ by working with our fixed set of gauge configurations generated with $L_s=8$
but varying the ``valence'' value of $L_s$ that appears in the propagators 
used to compute the ratio in Eq.~\ref{eq:mres_meas}.   The resulting $L_s$ dependence
of the resulting $m_{\rm res}$ can be viewed as a (likely good) approximation
to what would have been obtained if we also varied the $L_s$ appearing in the
fermion determinant.  This also can be viewed as a self-consistent study 
the properties of the 5-dimensional transfer matrix defined on gauge configurations
generated with this fixed, $L_s=8$ fermion action.

Next we examine the degree to which Eq.~\ref{eq:spec_density_2} describes the
dependence of $m_{\rm res}$ on $L_s$ in our simulations.  Here we rewrite that equation in the form:
\begin{equation}
m_{\rm res}(L_s) \simeq \left( c_1 e^{-\lambda_c L_s} + c_2 \right) \frac{1}{L_s}.
\label{eq:mresLs}
\end{equation}
Recall that the first term is expected to come from extended states with
eigenvalues $\lambda$ near the mobility edge: 
$\lambda_c \le \lambda \le \lambda_c + \frac{1}{L_s}$.  This term is exponentially 
suppressed for large $L_s$ and thus can be easily reduced by increasing $L_s$.  
The second term arises from low-lying localized states with 
$0 \le \lambda \le \frac{1}{L_s}$.  Here $c_2$ is proportional to the density of 
near-zero modes, $\rho(0)$. We require a demonstrably non-zero mobility edge 
$\lambda_c$ for safe QCD simulations with chiral formulations.  It is acceptable 
to have a significant non-exponential component in $m_{\rm res}$.  However, the
presence of such a term makes it difficult to significantly suppress chiral symmetry
breaking effects by simply increasing $L_s$. 

\begin{figure}
\includegraphics[width=\textwidth]{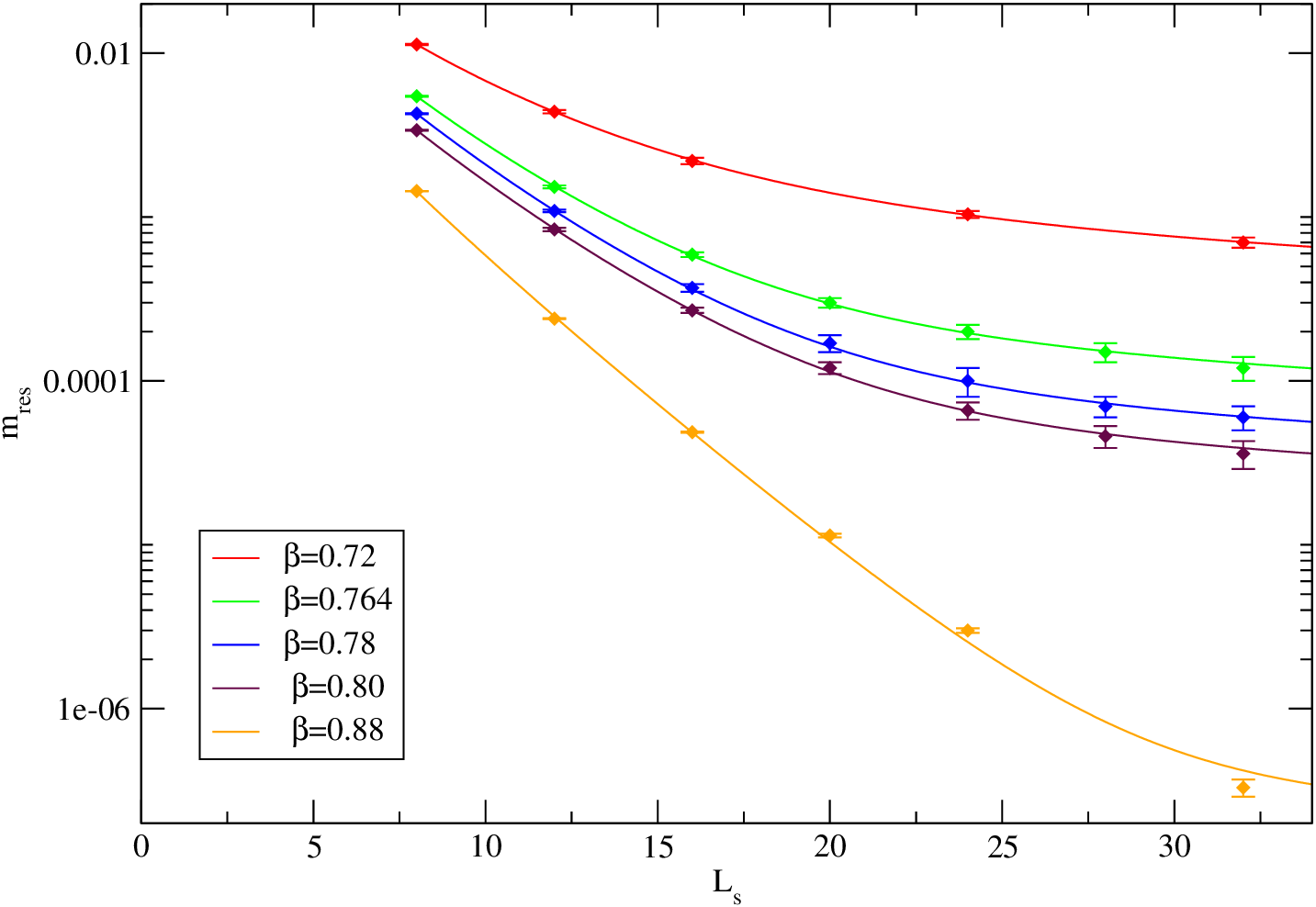}
\caption{\label{figMresDBW2}
Valence $m_{\rm res}(L_s)$ dependence of partially quenched DWF with the DBW2
gauge action as a function of $L_s$, using $L_s=8$ and the indicated
$\beta$ values on a $16^3\times 32$ volume for the ensembles. The sea and valence quark masses
are 0.04 throughout.
}
\end{figure}

\begin{figure}
\includegraphics[width=\textwidth]{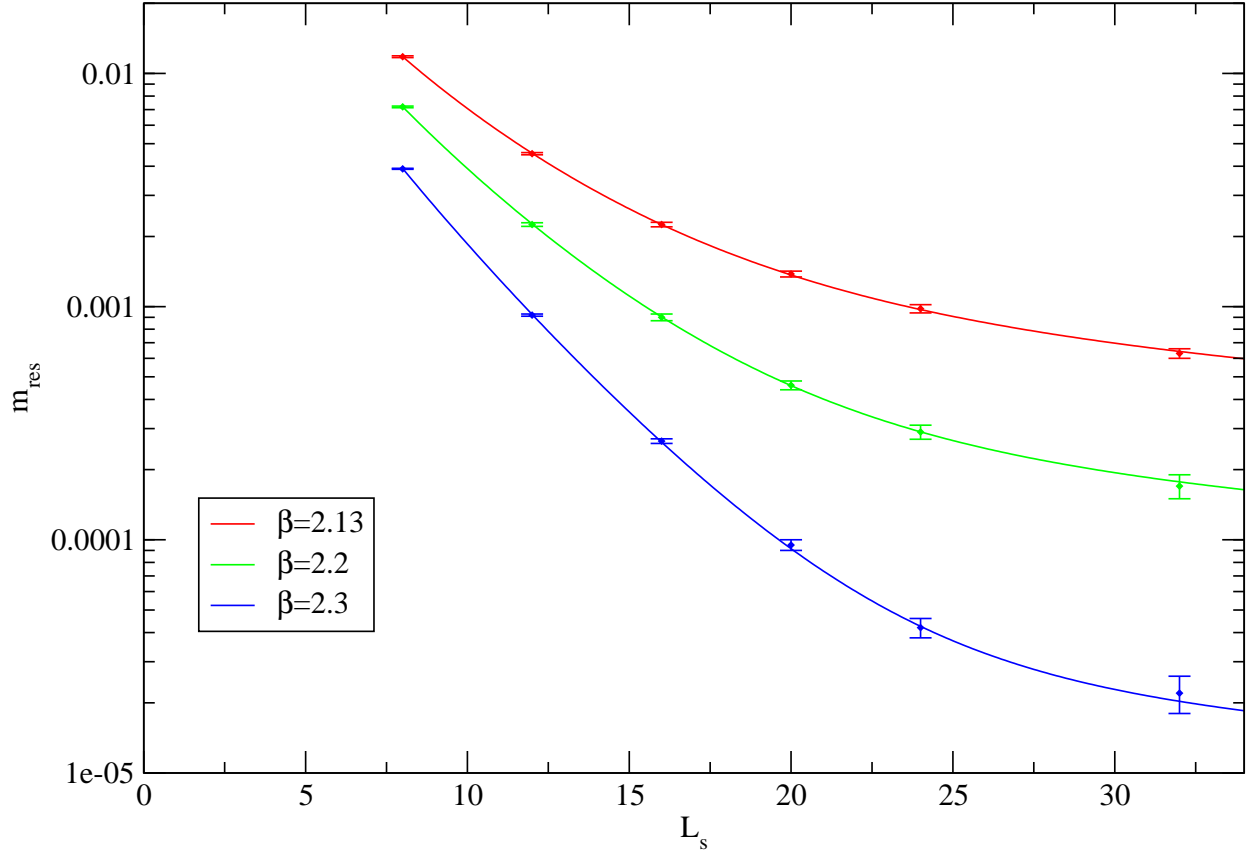}
\caption{\label{figMresIwasaki}
Valence $m_{\rm res}(L_s)$ dependence of partially quenched DWF with the 
Iwasaki gauge action as a function of $L_s$, using $L_s=8$ and the indicated
$\beta$ values on a $16^3\times 32$ volume for the ensembles. The sea and valence quark masses
are 0.04 throughout.
}
\end{figure}

We show fits to the functional form in Eq.~\ref{eq:mresLs} for  degenerate three flavor
ensembles at several gauge couplings for the Iwasaki and DBW2 gauge actions in 
Figures~\ref{figMresDBW2} and \ref{figMresIwasaki}.  The fitted parameters 
are listed in Table~\ref{TabFitMresIwasaki}.  While
the exponent $\lambda_c$ increases only slowly, the parameter $c_2$ falls by orders of magnitude
as the lattice spacing decreases. Interpolating 
between the values of lattice spacing given in these tables suggests the localized
near-zero modes are suppressed more by the DBW2 than the Iwasaki gauge 
action.  This behavior will be correlated with the rate of topology change 
discussed in the next section.

\begin{figure}
\includegraphics[width=\textwidth]{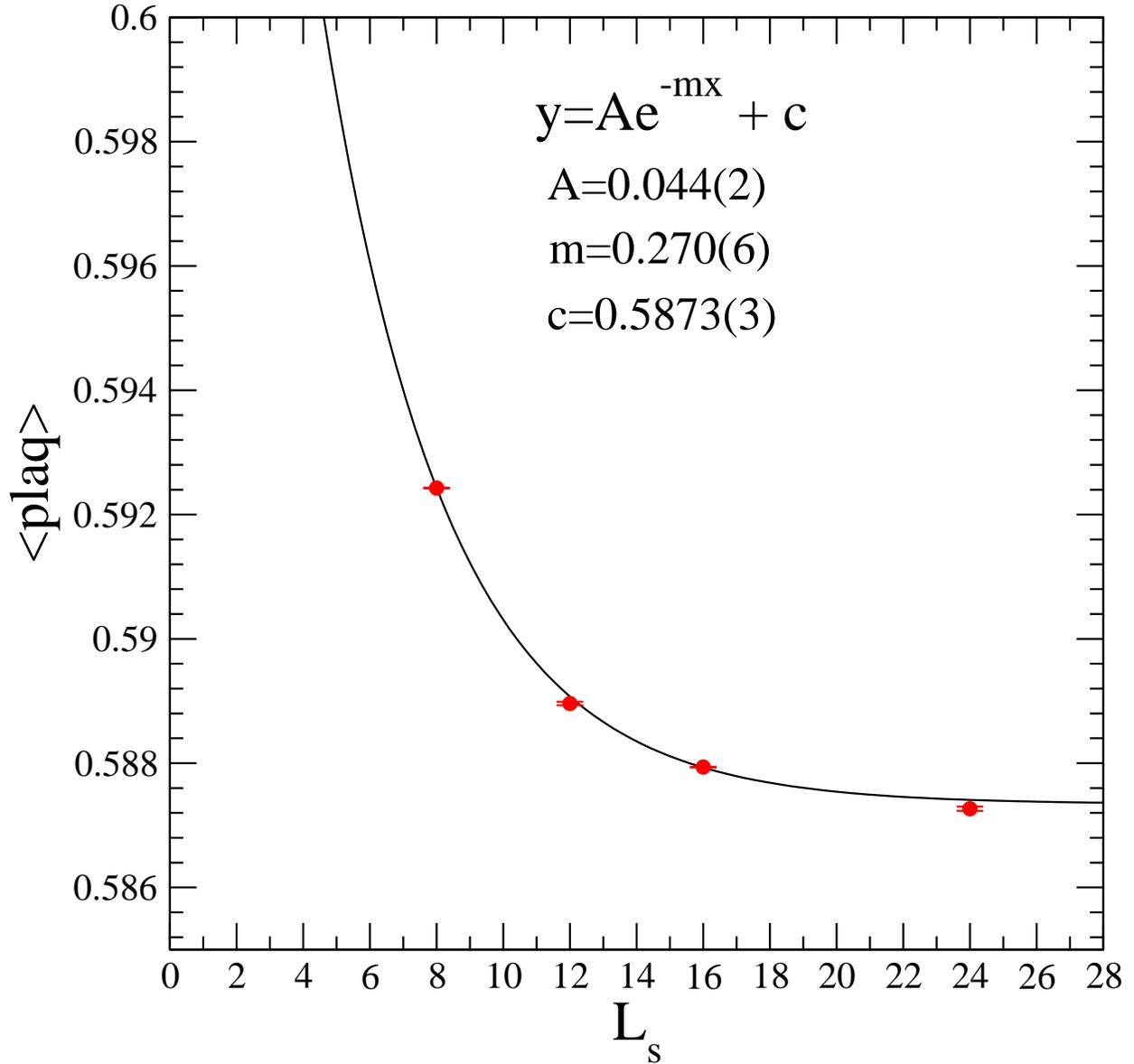}
\caption{\label{figPlaqLs}
Dependence of the plaquette as a function of the dynamical $L_s$, obtained
from low statistics runs. The dependence of this observable shows the expected $C + e^{-\alpha L_s}$
for an exponential approach to $L_s=\infty$, and thus demonstrates nicely the connection between
finite $L_s$ DWF simulations and those with exactly chiral fermions. 
}
\end{figure}

We note that while the residual mass term may be incorporated in a renormalized 
quark mass, one should extrapolate to $L_s = \infty$ to remove the residual 
chiral symmetry breaking effects of the $c_5$ term in Eq.~\ref{eq:eff_L}.  
However, even for finite $L_s$, such terms are expected to be very small compared
with even the errors obtainable with non-perturbative improvement techniques for
standard clover fermions, and thus can be treated as zero. Experience from clover 
fermions suggests that this level of error in the coefficient $c_5$
will certainly leave the results 
unmodified in practice, and that the unsuppressed $O(a^2)$ errors will be 
dominant in a continuum extrapolation. 
Figure~\ref{figPlaqLs} displays the dependence of the simplest of observables, the plaquette,
on $L_s$ for the sea quark content. It can be seen to be very weak (particularly
for $L_s\ge 16$) and consistent with an exponential approach to the $L_s\to \infty$ limit,
and is further evidence that all corrections to this limit are exponentially small.

\section{Localized Modes and Topology Change}

\label{secTopology}

Topology change, as defined by the index of an overlap
operator will, be accompanied by a change of sign of a mode of $H_W(-M_5)$
in the region $-2<-M_5<0$. There will be an instant $\tau = \tau_0(M_5)$ in the 
molecular dynamics
time at which $H_W(-M_5)$ has an exact zero mode for any appropriate
choice of $M_5$. Such a level crossing leaves the topological 
index of an overlap operator based on this kernel indefinite, depending on whether 
$-M_5$ is placed to the left or right of the level crossing, and it is reasonable
to conjecture that
low modes of $H_W(-M_5)$ play a significant role in topology change in
dynamical simulations with current algorithms.

In this section we shall present results for the topological charge time history
in our ensembles, and consider the relationship between tunneling rates,
the eigenmode spectrum, and details of the algorithm and action.

\begin{figure}
\includegraphics[width=\textwidth]{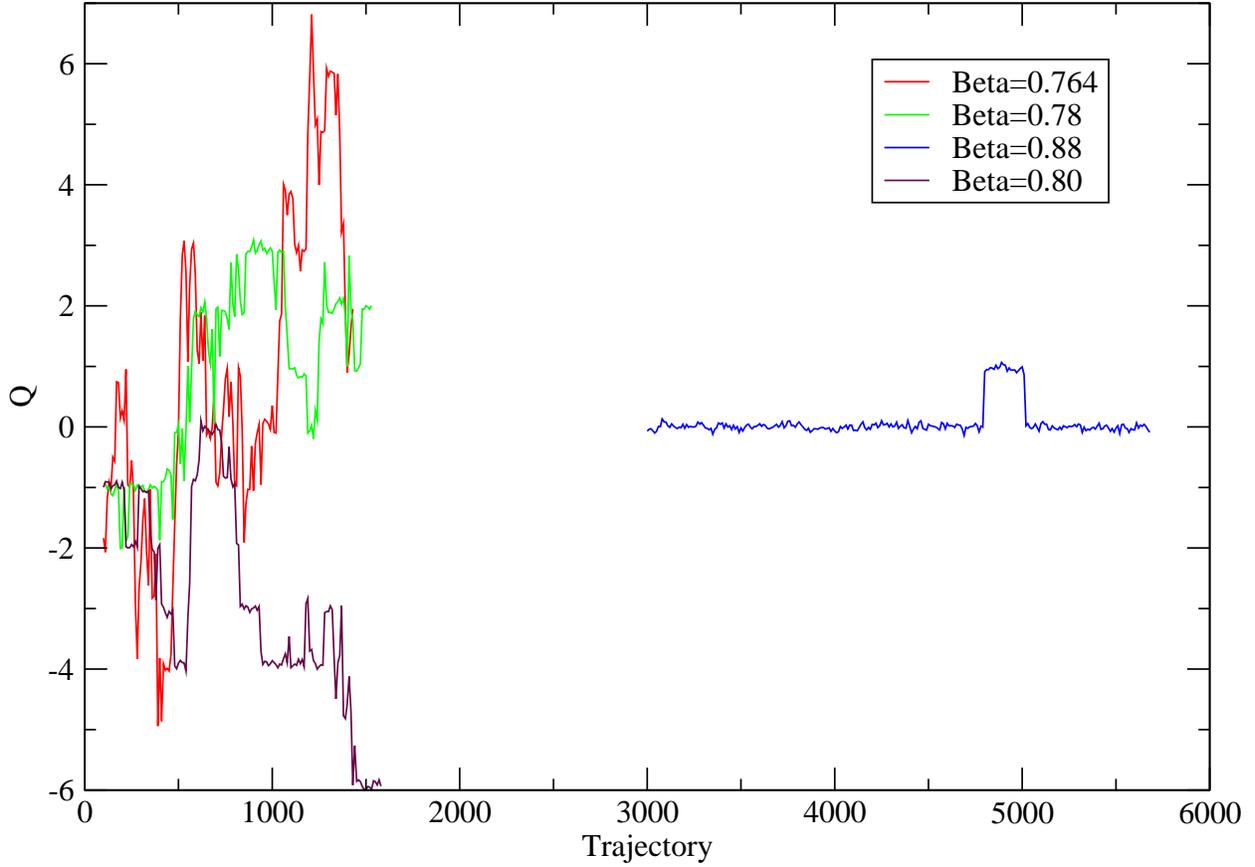}
\caption{\label{figTopoDBW2}
The Monte Carlo trajectory history for global topological charge with the DBW2 gauge action
at couplings $\beta=0.72,0.764,0.78,0.8$. The $\beta=0.88$ history is offset by 3000 and displays
a clear loss of topology tunneling. With a reasonable computational effort, a simulation
with the current algorithm and these parameters would be practically non-ergodic.
}
\end{figure}

\begin{figure}
\includegraphics[width=\textwidth]{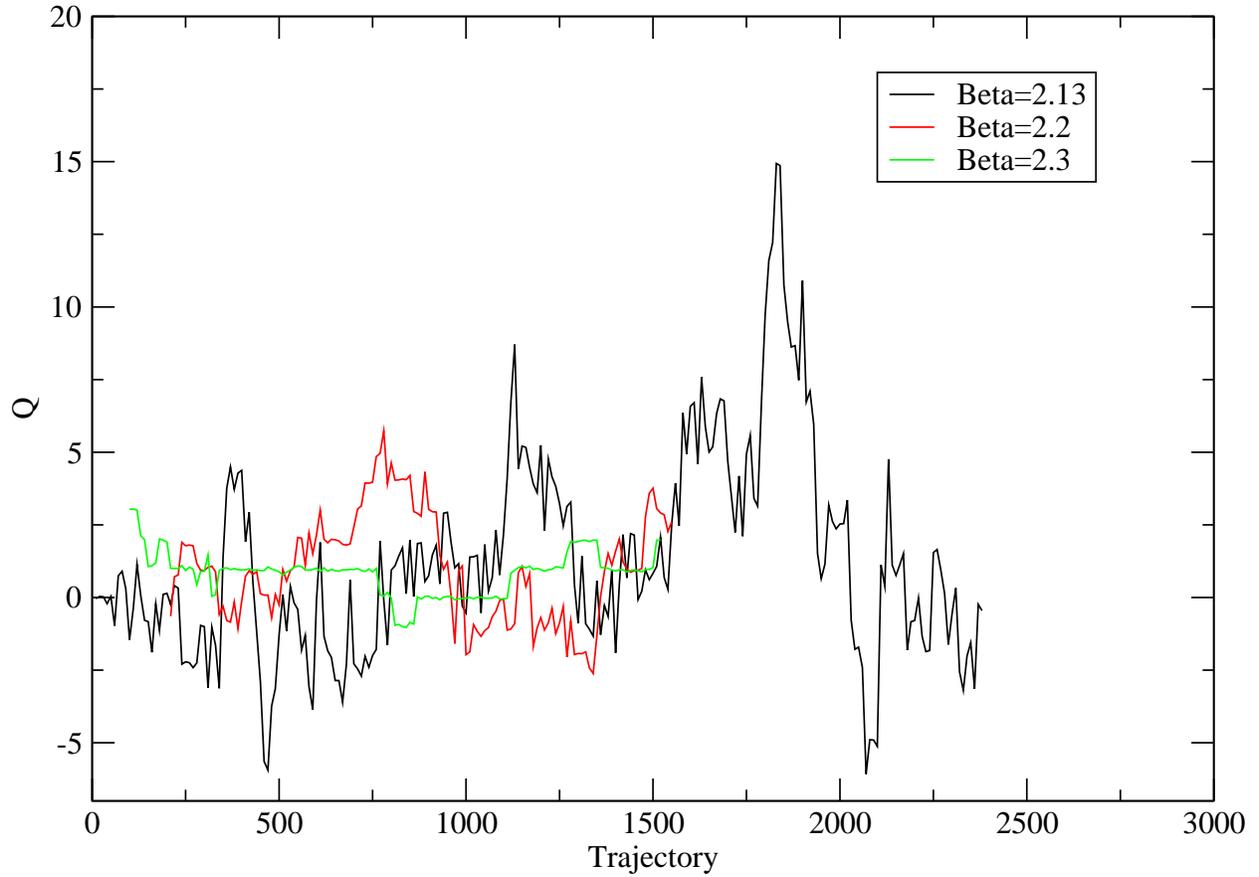}
\caption{\label{figTopoIwasaki}
The Monte Carlo trajectory history for global topological charge with the Iwasaki gauge action
at couplings $\beta=2.13,2.2,2.3$. The tunneling rate remains acceptable and leaves headroom for taking
a continuum limit with acceptable topological sampling without any further algorithmic improvement.
}
\end{figure}

In Figures~\ref{figTopoDBW2} and 
\ref{figTopoIwasaki} we display the topological charge time histories for 
all our datasets measured using the same $O(a^2)$-improved definition of the topological charge density
as in references \cite{Aoki:2005ga,Antonio:2006px}, implemented using
the Columbia Physics System. It is certainly the case that
the tunneling rate is substantially lower in those datasets with lower spectral densities. These are also
the configurations with the finest lattice spacings. 

Sampling all topological sectors has proven problematic with RG improved gauge actions
near the continuum limit even in the quenched approximation;
some of the authors found previously that the DBW2 gauge action samples topological sectors
rather poorly at $a^{-1} = 3 {\rm GeV}$ in the quenched 
case \cite{Aoki:2002vt}, and this loss of tunneling is likely driven by the gauge action. 
Going nearer to the continuum limit yields
an increasing potential barrier between topological sectors, arising from the gauge action,
which prevents topology change in those algorithms that either use a local update, or, even worse,
Hamiltonian evolution to propose a next configuration.

Whenever the determinant of $H_W$ does not directly enter the probability weight, 
the spectral density of low modes 
of $H_W$ found for the resulting gauge configurations is principally determined by the degree of roughness 
admitted by the gauge action, likely corresponding to fluxon-like configurations 
\cite{Edwards:1998sh,Berruto:2000fx}.
We conjecture that for domain wall fermions at practical values for $L_s$ the tunneling rate
is controlled by this low mode density.

\subsection{Tunneling with dynamical overlap fermions}

Additional problems have been seen with the introduction of exactly chiral fermions in
dynamical simulations. Exact implementations of the overlap operator
introduce a discontinuous step in the action when a low eigenvalue of $H_W$ changes sign
(and hence when the topological index of the configuration changes).
This infinitely narrow step in the action is guaranteed to be 
unresolved by numerical integration for any non-zero
timestep, and leads to a high probability of rejection of the proposed configuration. 
The reflection/refraction algorithm 
\cite{Egri:2005cx,Fodor:2003bh} is one sensible, but expensive, response to this. 

Another response has been to develop actions which 
suppress low modes of $H_W$ with pseudofermion determinant estimation
\cite{Vranas:1999rz,Izubuchi:2002pq,Fukaya:2006vs}, and so suppress topology change.
While many hadronic quantities
are not likely to be sensitive to global topology (especially in the limit of large volume), 
there remains a serious concern that a modification which suppresses the
change of global topology may also interfere with the creation of a proper
distribution of local topological fluctuations.  In the language of instantons,
one should demonstrate that such an algorithm leads to a physical density of
instanton-anti-instanton pairs -- a quantity that may be important for QCD
even when the global topology is fixed.
However, we have poor tools for knowing whether this
globally non-ergodic algorithm is correctly
sampling local topological fluctuations, given that we have thrown away our
best diagnostic (and indeed the indicator that originally flagged the problem), 
i.e. global topology change.

So far with DWF we have obtained acceptable tunneling rates, and this is likely
because the response to the lowest eigenvalues of the $\tanh$ approximation to the sign function,
Eq.~\ref{eq:Doverlap}, is spread out at the level of our integration timestep. 
More specifically the determinant whose value is estimated contains a contribution
$\tanh L_s \tanh^{-1} \lambda$, in place of ${\rm sgn} \lambda$, and the 
transition has a width of $O(\frac{1}{L_s})$. If we consider the low mode to be evolving 
in molecular dynamics time, with timestep $\delta\tau$ and at a rate $\dot{\lambda}$, then the
molecular dynamics integral will be accurate provided 

\begin{equation} 
\label{eq:dtbound}
\delta\tau \ll \frac{1}{L_s\dot{\lambda}}.
\end{equation}

There will be no practical impact on the tunneling rate from increasing $L_s$ while the condition 
Eq.~\ref{eq:dtbound} is held, and the problem may, in fact, never arise for any practical DWF simulation. 
In practice, the tunneling rate is determined by other effects, such as the low mode density determined
by roughness admitted by the gauge action. We have also found a significant improvement in the tunneling
rate, subsequent to this simulation arising from improving
the quality of our stochastic estimate of the fermion force \cite{Allton:2007hx}, where
the use of a single stochastic estimate for the ratio of Pauli-Villars and light mass determinants
was found to allow increased molecular dynamics step size and to increase the tunneling rate.

This also suggests 
a third possible response to this problem that arises in simulations with better 
approximations to the overlap operator than our
DWF simulations. That is to relax the approximation in the molecular dynamics 
evolution such that Eq.~\ref{eq:dtbound}
is satisfied, while keeping the approximation accurate for the accept/reject
step. This differential treatment of Metropolis and 
molecular dynamics steps is not possible with DWF simulations due to the fully five-dimensional
pseudofermion fields, but it is possible for many approaches to the overlap operator.

\section{Conclusion}
\label{secConclusion}

When constrained to keep $L_s$ affordable, 
there is some degree of trade-off between the exactness of chirality and 
the thoroughness of topological sampling with domain wall fermions. 
Both are determined by the density of modes in the region where the $\tanh$ approximation
is inaccurate, and this density is
controlled by either the choice of the gauge action, or the inclusion of additional
unphysical terms in the action.

One might pause for a moment to ponder where to compromise. 
Using computer precision, ({\it e.g.} single precision) 
as the standard for achieving chiral symmetry invokes computer technology rather than physics for an 
answer to this important question.
The authors' conclusion from this detailed study 
is that we should require \emph{adequate} chirality, 
and by that we mean that the average symmetry breaking
should not compromise the attractive simplification of electroweak current structure and 
off-shell $O(a)$ improvement for renormalization.  

We have demonstrated in this paper that we can obtain $m_{\rm res}\simeq O(10^{-3})$ with both the DBW2 and Iwasaki
gauge actions in the region of $a^{-1} \simeq 1.6 {\rm GeV}$.  We have observed that dynamical simulations with
the DBW2 gauge action rapidly lose topological tunneling as $\beta$ is increased. 
In contrast, headroom is left for going
to weaker coupling, while maintaining good topological sampling, with the Iwasaki gauge action. This approach
is certainly at best a stay of execution, as topological tunneling will be lost on sufficiently fine lattices.
These are inaccessible with our current computers and
our conclusion is thus clear - the RBC and UKQCD collaborations are running \cite{Allton:2007hx}
dynamical domain wall fermions with the Iwasaki gauge action and $L_s \ge 16$. 
We started from $a^{-1}\simeq 1.6{\rm GeV}$ and thus we are able to
maintain both excellent topological sampling and acceptable chiral symmetry violation.  

\section*{Acknowledgements}
 
 We thank Dong Chen, Calin Cristian, Zhihua Dong, Alan Gara, Andrew
 Jackson, Changhoan Kim, Ludmila Levkova, Xiaodong Liao, Guofeng
 Liu, Konstantin Petrov and Tilo Wettig for developing with us the QCDOC
 machine and its software. This development and the resulting computer
 equipment used in this calculation were funded by the U.S.\ DOE grant
 DE-FG02-92ER40699, PPARC JIF grant PPA/J/S/1998/00756 and by RIKEN. This work
 was supported by DOE grants DE-FG02-92ER40699 and DE-AC02-98CH10886 and PPARC grants
 PPA/G/O/2002/00465, PP/D000238/1 and PP/C504386/1. AH is supported by the UK Royal Society.
 We thank BNL, EPCC, RIKEN, and the U.S.\ DOE for
 supporting the computing facilities essential for the completion of this work.

\appendix

\section{Domain wall fermion transfer matrix}
\label{sec:transfer_matrix}

Here we collect some useful formulae that describe the transfer matrix
formalism as applied to domain wall fermions by Furman and 
Shamir~\cite{Furman:1994ky}.  While we do not present a derivation of 
these formulae, they follow reasonably directly from the results
presented by Furman and Shamir, Narayanan and Neuberger~\cite{Narayanan:1994sk} 
and L\"uscher~\cite{Luscher:1976ms}.

We begin with the five-dimensional domain wall fermion Dirac operator 
defined in Eqs.~\ref{eq:D_dwf}, \ref{eq:D_parallel} and \ref{eq:D_perp}.  
(This definition of the domain wall fermion operator is consistent with 
our earlier papers, but differs by a hermitian conjugate from the 
original notation of Furman and Shamir.)  We also identify two important 
4-dimensional operators which play a central role in L\"uscher's original 
transfer matrix construction~\cite{Luscher:1976ms}:
\begin{eqnarray}
B_{x,x'} &=& (5-M_5)\delta_{x,x'} 
       - \frac{1}{2}\sum_\mu\Bigl[\delta_{x+\mu,x'}U_{x ,\mu} 
                                 +\delta_{x,x'+\mu}U_{x',\mu}^\dagger  \Bigr] 
\label{eq:B_def} \\
C_{x,x'} &=& 
         \frac{1}{2}\sum_\mu\Bigl[\delta_{x+\mu,x'}U_{x ,\mu} 
                                - \delta_{x,x'+\mu}U_{x',\mu}^\dagger \Bigr]\sigma^\mu.
\label{eq:C_def}
\end{eqnarray}
Here we have adopted a spinor basis in which $\gamma^5$ is diagonal.  In this basis 
the $\gamma$ matrices can be written
\begin{equation}
\gamma^\mu = \left ( \begin{array}{cc} 0 &  \sigma^\mu \\
                   (\sigma^\mu)^\dagger &  0 \end{array} \right) \quad
\gamma^5 = \left ( \begin{array}{cc} 1 &  0 \\
                                     0 &  -1 \end{array} \right)\label{eq:gamma_def}
\end{equation}
and the four $2 \times 2$ matrices $\sigma^\mu$ appearing in both 
Eqs.~\ref{eq:C_def} and \ref{eq:gamma_def} are given by 
$\sigma^\mu = (i,\vec\sigma)$, where $\vec \sigma$ represents the three 
standard Pauli matrices and $\frac{1}{2}(1\pm\gamma^5)$ projects onto
right-/left-handed states.  Thus, the matrices $B$ and $C$ have a 
$2\times 2$-dimensional spin structure and appear directly in the matrix 
$D^\parallel_{x,x^\prime}(M_5)$ of Eq.~\ref{eq:D_parallel}:
\begin{eqnarray}
D^\parallel_{x,x^\prime}(M_5) 
&=& I - \left(\begin{array}{cc} B         & C \\
                                C^\dagger & B 
                                \end{array}\right).
\label{eq:D_parallel_2}
\end{eqnarray}

Next we need to introduce three pairs of related fermion fields.  The first two
pairs are the 5-dimensional Grassmann variables $\Psi_{x,s}$ and $\overline\Psi_{x,s}$
and the corresponding Fock-space operators $\hat \psi_x$ and $\hat{\overline\psi}_x$.  
The field operators $\hat \psi_x$ and $\hat{\overline\psi}_x$ can be written in
terms of two-component field operators $\hat c_x$ and $\hat d_x$ and their
hermitian conjugates:
\begin{eqnarray}
\hat\psi         &=& \left ( \begin{array}{c} B^{-\frac{1}{2}}(\hat d^\dagger)^t \\
                                              B^{-\frac{1}{2}}\hat c \end{array}\right) 
\label{eq:c_d_def0}\\
\hat{\overline\psi} &=& \left ( \begin{array}{cc} -\hat d\,^t B^{-\frac{1}{2}}, \quad 
                         \hat c^\dagger B^{-\frac{1}{2}} \end{array}\right ). 
\label{eq:c_d_def1}
\end{eqnarray}
Here the operators $\hat c$ and $\hat d$ both act on vectors in the Fock space and 
carry space-time, color and spinor indices.  As Fock space operators, they 
can be thought of as $2^{12 L^3 L_t}$ dimensional square matrices if we are 
working with a $L^3 \times L_t$ space-time lattice.  We treat their space-time, 
color and spin indices as the index on a $6 L^3 L_t \times 1$ column vector.  
It is on this vector that the $6 L^3 L_t \times 6 L^3 L_t $ matrix 
$B^{-\frac{1}{2}}$ acts in Eqs.~\ref{eq:c_d_def0} and \ref{eq:c_d_def1}.  
We use the conventions that the hermitian conjugate, represented by the 
superscript $\dagger$ takes the hermitian conjugate of both the Fock-space 
operator and the space-time-flavor-spin matrix.  The transpose, represented 
by the superscript $t$, acts only on the space-time-flavor-spin matrix.

Finally we introduce two further operators:
\begin{eqnarray}
\hat R(\hat c, \hat d) &=& \exp\Bigl\{\hat d^t(B^{-\frac{1}{2}}C B^{-\frac{1}{2}})\hat c\Bigr\} \\
\hat W(\hat c^\dagger, \hat d^\dagger, \hat c, \hat d) 
                  &=& \exp\Bigl\{-\hat c^\dagger \ln(B) \hat c -\hat d^\dagger \ln(B) \hat d \Bigr\}.
\end{eqnarray}
Using these operators we can then connect the Feynman path integral, which 
provides the original definition of domain wall fermion lattice theory,
with a Fock-space operator expression.  For simplicity, we consider an
example of the 5-dimensional fermion propagator, because a general fermionic
Green's function follows exactly the same pattern:
\begin{eqnarray}
\int \prod_{x',s'} d[\Psi_{x',s'}] d[\overline\Psi_{x',s'}] e^{\overline\Psi D_{\rm DWF} \Psi} \;
                          \Psi_{x_1,s_1}\overline\Psi_{x_2,s_2} && \nonumber \\
&& \hskip -2.0in =\prod_{s=0}^{L_s-1}\Biggl\{ 
\prod_x \int d\overline \Psi_{x,s} d \Psi_{x,s}  \exp \Bigl[
\overline\Psi^R_s \Psi^L_{s+1} + \overline\Psi^L_{s+1} \Psi^R_s 
- \overline\Psi^L_s B \Psi^R_s - \overline \Psi^R_s B \Psi^L_s
\nonumber \\
&&\hskip -1.5in - \overline\Psi^R_s C^\dagger \Psi^R_s + \overline \Psi^L_s C \Psi^L_s \Bigr] \Biggr\}
\Psi_{x_1,s_1}\overline\Psi_{x_2,s_2} \label{eq:ord_prod_step} \\
&& \hskip -2.0in = (\det B)^{2L_s}\;\mbox{tr}\Biggl\{
                       \Bigl(\hat W^\frac{1}{2}\hat R \hat R^\dagger \hat W^\frac{1}{2}\Bigr)^{s_1}
                             \hat W^\frac{1}{2}\hat R \hat\psi_{x_1} \hat R^\dagger \hat W^\frac{1}{2}
                       \Bigl(\hat W^\frac{1}{2}\hat R \hat R^\dagger \hat W^\frac{1}{2}\Bigr)^{s_2-s_1-1}
\nonumber \\
&& \hskip -1.5in             \hat W^\frac{1}{2}\hat R \hat{\overline\psi}_{x_2} \hat R^\dagger \hat W^\frac{1}{2}
                       \Bigl(\hat W^\frac{1}{2}\hat R \hat R^\dagger \hat W^\frac{1}{2}\Bigr)^{L_s-s_2-1}
                               {\cal{O}}(m_f)        \Biggr\}
\label{eq:ord_prod_gen}
\end{eqnarray}
where
\begin{equation}
         {\cal{O}}(m_f) = \prod_\alpha\Bigl\{(\hat c_\alpha \hat c_\alpha^\dagger + m_f  \hat c_\alpha^\dagger \hat c_\alpha)
                                     (\hat d_\alpha \hat d_\alpha^\dagger + m_f  \hat d_\alpha^\dagger \hat d_\alpha)
                              \Bigr\}
\label{eq:Omf_def}
\end{equation}
and $\alpha$ represents space-time, spin and color indices.  While a complete derivation 
of the result in Eq.~\ref{eq:ord_prod_gen} is beyond the scope of this paper, one of the
intermediate steps is shown in Eq.~\ref{eq:ord_prod_step}.  In that equation the Grassmann
fields $\Psi$ and $\overline{\Psi}$ have been expressed in terms of their chiral components
according to:
\begin{eqnarray}
         \Psi = \left(\begin{array}{c} \Psi^R \\ \Psi^L \end{array}\right) \quad
\overline\Psi = \left(\begin{array}{cc} \overline{\Psi}^L & \overline{\Psi}^R \end{array}\right).
\end{eqnarray}
A complete derivation of Eq.~\ref{eq:ord_prod_gen} can be constructed from 
Refs.~\cite{Luscher:1976ms,Narayanan:1994sk,Furman:1994ky}.  Note, the Grassmann 
fields $\Psi^L_{L_s}$ and $\overline{\Psi}^L_{L_s}$ appear in the integrand in 
Eq.~\ref{eq:ord_prod_step} but are not among the variables of integration.  Instead, 
these fields are to be evaluated as $-m_f$ times the corresponding fields at $s=0$: 
$-m_f \Psi^L_{s=0}$ and $-m_f \overline{\Psi}^L_{s=0}$, respectively.

Following Furman and Shamir, we identify the Fock-space transfer matrix $T$ as
\begin{eqnarray}
T  &=& (\det B) \hat W^\frac{1}{2}\hat R \hat R^\dagger \hat W^\frac{1}{2} = e^{-\hat a^\dagger H_T \hat a} 
\label{eq:T_def}\\
\mbox{where}\quad \hat a &=& \left ( \begin{array}{c} (\hat d^\dagger)^t \\ \hat c \end{array}     \right),  
\quad \hat a^\dagger      =  \left ( \begin{array}{cc} \hat d\,^t, \quad \hat c^\dagger \end{array}\right)
\quad \mbox{and} \label{eq:a_def} \\
e^{-H_T}&=& \left (\begin{array}{cc}  B+C B^{-1}C^\dagger & C B^{-1}   \\             
                                      B^{-1}C^\dagger     & B^{-1}\end{array} \right).
\label{eq:H_def}
\end{eqnarray}

The unusual arrangement of the operators $\hat\psi_{x_1}$ and 
$\hat{\overline\psi}_{x_2}^\dagger$ and two missing factors of $T$ in 
Eq.~\ref{eq:ord_prod_gen} can be understood if, following Furman and Shamir, 
we recognize that the upper component of $\hat\psi$ and the lower component 
of $\hat{\overline\psi}$ commute with $\hat R^\dagger$, while the lower 
component of $\hat\psi$ and the upper component of $\hat{\overline\psi}$ 
commute with $\hat R$.  This allows the operators $\hat\psi$ and 
$\hat{\overline\psi}$ to be extracted from between the factors 
$\hat W^\frac{1}{2}\hat R$ and $\hat R^\dagger \hat W^\frac{1}{2}$, so that the 
two missing factors of $T$ can be assembled.  (Note, the operators $\hat\psi$ 
and $\hat{\overline\psi}$ do not commute with the operator $\hat W^\frac{1}{2}$, 
but instead have the non-local factor of $B^{-\frac{1}{2}}$ appearing in 
Eqs.~\ref{eq:c_d_def0} and \ref{eq:c_d_def1} removed by the proposed 
interchange).

We will demonstrate this re-arrangement of operators by working out two
examples.  In the first, we translate the expectation value
$\langle\overline q(x) q(x)\rangle$ (the chiral condensate) into this 
transfer matrix language.  Recall that the physical, 4-dimensional 
Grassmann fields are defined by:
\begin{eqnarray}
q(x)           &=& P_L\Psi_{x,0} +P_R\Psi_{x,L_s-1} \label{eq:q_def} \\
\overline q(x) &=& \overline \Psi_{x,0} P_R + \overline\Psi_{x,L_s-1}P_L,
\label{eq:qbar_def}
\end{eqnarray}
where we have used $P_{R/L} = (1 \pm \gamma^5)/2$.   If these expressions 
are substituted into Eq.~\ref{eq:ord_prod_gen}, we obtain:
\begin{eqnarray}
\langle \overline q(x) q(x) \rangle &=& -(\det B)^{L_s} \mbox{tr} 
  \Biggl\{\hat W ^\frac{1}{2}\hat R 
   (P_L\hat \psi_{x})_\alpha \hat R^\dagger \hat W^\frac{1}{2} 
\label{eq:qbq} \\
    &&\quad\quad      T^{L_s-2} \hat W ^\frac{1}{2}\hat R 
   (\hat{\overline\psi}_xP_L)_\alpha\hat 
          R^\dagger \hat W ^\frac{1}{2} {\cal{O}}(m_f) \Biggr\}
\nonumber \\
&+& (\det B)^{L_s} \mbox{tr} 
  \Biggl\{\hat W ^\frac{1}{2}\hat R 
   (\hat{\overline\psi}P_R)_\alpha \hat R^\dagger \hat W^\frac{1}{2}
\nonumber \\
    &&\quad\quad      T^{L_s-2} \hat W ^\frac{1}{2}\hat R 
   (P_R\hat \psi_x)_\alpha\hat 
          R^\dagger \hat W ^\frac{1}{2} {\cal{O}}(m_f) \Biggr\}.
\nonumber
\end{eqnarray}
The final step involves commuting the $P_L\hat\psi$ and $\hat{\overline\psi}P_R$ 
past the factor $\hat W^\frac{1}{2} \hat R$ to stand on the far left.  Similarly, 
the operators $P_R\hat\psi$ and $\hat{\overline\psi}P_L$ on the right side can 
be moved past the factor $\hat R^\dagger \hat W^\frac{1}{2}$ to stand directly to the 
left of the operator ${\cal{O}}(m_f)$.  The resulting expression can be written simply 
in terms of the operators $\hat a$ and $\hat a^\dagger$:
\begin{eqnarray}
\langle \overline q(x) q(x) \rangle &=& -(\det B)^{L_s} \mbox{tr} 
  \Biggl\{ \Bigl[(P_L\hat a_{x})_\alpha T^{L_s} (\hat a^\dagger_xP_L)_\alpha 
\label{eq:qbq2} \\ 
&&\quad \quad \quad \quad + (\hat a_{x}^\dagger P_R)_\alpha  T^{L_s} 
            (P_R\hat a_x)_\alpha \Bigr] {\cal{O}}(m_f) \Biggr\}.
\nonumber
\end{eqnarray}

As a second example, consider the Green's function 
$\langle P_L q(x)\;J^b_{5q}(z)\;\overline q(y)P_L \rangle$ containing the midpoint
operator $J^a_{5q}$ that occurs in the divergence of the 5-dimensional flavor
non-singlet axial current introduced by Furman and Shamir~\cite{Furman:1994ky}:
\begin{equation}
J^b_{5q}(x) = -\overline\Psi(x,L_s/2-1)P_L t^b\Psi(x,L_s/2) 
                 + \overline\Psi(x,L_s/2)P_R t^b \Psi(x,L_s/2-1).
\label{eq:j5q_def}
\end{equation}
Here $b$ is a flavor index and $t^b$ is a flavor generator.  

Using Eq.~\ref{eq:ord_prod_gen} for this product of four Grassmann fields we
find:
\begin{eqnarray}
\langle P_L q(x)\;J^b_{5q}(z)\;\overline q(y)P_L \rangle &=& -(\det B)^{L_s} \mbox{tr} 
  \Biggl\{\hat W ^\frac{1}{2}\hat R 
   (P_L\hat \psi_{x}) \hat R^\dagger \hat W^\frac{1}{2} T^{L_s/s-2}
\label{eq:5q} \\
    &&\Bigl[-\hat W^\frac{1}{2} \hat R 
              (\hat{\overline\psi}_z P_L t^b)_\alpha \hat R^\dagger \hat W^\frac{1}{2}
                  \hat W^\frac{1}{2} \hat R (P_L\hat\psi_z)_\alpha \hat R^\dagger \hat W^\frac{1}{2}
\nonumber \\
    &&\quad  -\hat W^\frac{1}{2} \hat R 
              (t^b P_R\hat\psi_z)_\alpha \hat R^\dagger \hat W^\frac{1}{2}
                  \hat W^\frac{1}{2} \hat R (\hat{\overline\psi}_zP_R)_\alpha \hat R^\dagger \hat W^\frac{1}{2}
\Bigr] 
\nonumber \\
&&\quad\quad T^{L_s/s-2} \hat W ^\frac{1}{2}\hat R (\hat{\overline\psi}_y P_R)\hat 
          R^\dagger \hat W ^\frac{1}{2} {\cal{O}}(m_f) \Biggr\}.
\nonumber
\end{eqnarray}
Performing the same operator manipulations as used to obtain Eq.~\ref{eq:qbq}
we find:
\begin{eqnarray}
\langle P_L q(x)\;J^b_{5q}(z)\;\overline q(y)P_L \rangle 
  &=& (\det B)^{L_s}\mbox{tr} \Bigl\{ P_L \hat a_x T^\frac{L_s}{2}
\hat a^\dagger_z t^b \hat a_z T^\frac{L_s}{2}\hat a^\dagger_y P_L {\cal{O}}(m_f) \Bigl\}.
\label{eq:j5q_transf}
\end{eqnarray}
Here the unit displacement in $s$ between the two factors in $J^a_{5q}$
has disappeared leaving a very symmetrical expression.  The missing $\gamma^5$
matrix, which one would expect in an axial matrix element, has its origins in
the minus sign that is present in Eq.~\ref{eq:c_d_def1} but absent in 
Eq.~\ref{eq:a_def}.  This factor could be restored if we introduced 
$\hat{\overline a}_x = \hat a_x^\dagger \gamma^5$, a 5-dimensional analogue 
of the role played by $\gamma^0$ in the standard treatment of the 4-dimensional
Dirac operator.

As a final topic for this appendix, we will examine the effects of the 
matrix $T^{L_s}$ in Eq.~\ref{eq:qbq}.  Again following Furman and Shamir,
we introduce the eigenvectors and eigenvalues of the operator $H_T$ defined
in Eq.~\ref{eq:H_def}:
\begin{eqnarray}
H_T \phi^+_{k^+} = E^+_{k^+}\phi^+_{k^+} \quad 1 \le k^+ \le N^+ \\
H_T \phi^-_{k^-} = -E^-_{k^-}\phi^-_{k^-} \quad 1 \le k^- \le N^-
\end{eqnarray}
where $E^\mp_n \ge 0$, and write the operator $\hat a_x$ in terms of this basis,
filling the Dirac sea in the usual way :
\begin{eqnarray}
\hat a_x = \sum_{k+=1}^{N^+} \phi^+_{k^+}(x)\; \hat o_k 
                  + \sum_{k^-=1}^{N^-}  \phi^-_{k^-}(x)\; \hat p_{k^-}^\dagger.
\label{eq:a_expansion}
\end{eqnarray}
Let $|0_H\rangle$ be the Fock space state $|0_H\rangle$ which is the eigenstate 
of $T$ with the largest eigenvalue.  This state obeys:
\begin{equation}
\hat o_{k^+} |0_H\rangle = \hat p_{k^-} |0_H\rangle = 0
\end{equation}
for all $k^+$ and $k^-$ and 
\begin{equation}
T |0_H\rangle = \lambda_{\rm max} |0_H\rangle 
              = \exp\left\{\sum_{k^-=1}^{N^-} E^-_{k^-}\right\} |0_H\rangle.
\end{equation}
Finally we define the normalized operator $\hat T = T /\lambda_{\rm max}$ making
unity the largest eigenvalue of $\hat T$ and $|0_H\rangle$ the corresponding 
eigenstate.  The matrix $\hat T$ can be written:
\begin{eqnarray}
\hat T &=& T/\lambda_{\rm max} = \exp\left\{-\hat a ^\dagger H \hat a \right\}/\lambda_{\rm max} \\
       &=& \exp\left\{-\sum_{k^+=1}^{N^+} E^+_{k^+}\hat o_{k^+}^\dagger \hat o_{k^+} 
                      +\sum_{k^-=1}^{N^-} E^-_{k^-}\hat p_{k^-} \hat p_{k^-}^\dagger \right\}/\lambda_{\rm max} \\
       &=& \exp\left\{-\sum_{k^+=1}^{N^+} E^+_{k^+}\hat o_{k^+}^\dagger \hat o_{k^+} 
                      -\sum_{k^-=1}^{N^-} E^-_{k^-}\hat p_{k^-}^\dagger \hat p_{k^-} \right\}. 
\label{eq:T_diag}
\end{eqnarray}

Thus, in the limit of large $L_s$, $\hat T^{L_s}$ becomes the projection 
operator onto the state $|0_H\rangle$, separating the right and left 
operators in an equation such as Eq.~\ref{eq:qbq} into two separate 
factors, linked by the operator ${\cal{O}}(m_f)$.  If we further take the limit 
$m_f=0$ and define a surface vacuum state $|0_S\rangle$, which is
annihilated by the operators $\hat c_x$ and $\hat d_x$,
\begin{equation}
\hat c_x |0_S\rangle = \hat d_x |0_S\rangle = 0,
\end{equation}
then the operator ${\cal{O}}(m_f=0)$, given in Eq.~\ref{eq:Omf_def}, reduces to the
projection operator $|0_S\rangle \langle 0_S|$ and the left- and right-handed 
sectors become completely independent matrix elements between $|0_S\rangle$ 
and $|0_H\rangle$.

This can be summarized by evaluating a general Green's function depending on 
the 4-dimensional fields $q$ and $\overline q$.  This is most easily
done if the fermion fields in the original Green's function are ordered with
the left-handed chiral fields ($P_L q$ and $\overline q P_R$) on the left 
and the right-handed chiral fields ($P_R q$ and $\overline q P_L$) on the 
right:
\begin{eqnarray}
\frac{1}{(\det B\; \lambda_{\rm max})^{L_s}} 
\left\langle {\cal O}_L[P_L q,\overline q P_R]\; 
              {\cal O}_R[P_R q,\overline q P_L]\right\rangle_{L_s}\hskip -2.5in && \\
&&=\frac{1}{(\lambda_{\rm max})^{L_s}} \;
{\rm tr}\left\{ {\cal O}_L[P_L\hat a, -\hat a^\dagger P_R] 
         T^{L_s}{\cal O}_R[P_R \hat a, \hat a^\dagger P_L] {\cal{O}}(m_f)\right\} \\
&&\hskip -0.1in \mathrel{\mathop\rightarrow_{L_s \to \infty}}
{\rm tr}\left\{ {\cal O}_L[P_L\hat a, -\hat a^\dagger P_R] 
         |0_H\rangle \langle 0_H| {\cal O}_R[P_R \hat a, \hat a^\dagger P_L] {\cal{O}}(m_f)\right\} \\
&&\hskip -0.1in \mathrel{\mathop\rightarrow_{m_f \to 0}}
\langle 0_S |{\cal O}_L[P_L\hat a, -\hat a^\dagger P_R] 
         |0_H\rangle \langle 0_H| {\cal O}_R[P_R \hat a, \hat a^\dagger P_L] | 0_S \rangle,
\label{eq:chiral_factorization}
\end{eqnarray}
where the quantities ${\cal O}_L$ and ${\cal O}_R$ are two ordered polynomials
in their two arguments.

Corrections to the limit $L_s \to \infty$ can be written by including 
eigenvectors of $\hat T$ with smaller eigenvalues.  Such corrections can
be easily obtained from Eq.~\ref{eq:T_diag}:
\begin{equation}
\hat T^{L_s} \approx |0_H\rangle \langle 0_H| 
                 +\sum_{k^+=1}^{N^+} e^{-E^+_{k^+}L_s}\hat o_{k^+}^\dagger |0_H\rangle \langle 0_H|\hat o_{k^+} 
                 +\sum_{k^-=1}^{N^-} e^{-E^-_{k^-}L_s}\hat p_{k^-}^\dagger |0_H\rangle \langle 0_H|\hat p_{k^-}
                 +\ldots.
\label{eq:T_corr}
\end{equation}

\bibliography{paper}


\begin{table}[hbt]
\caption{\label{tabParams} Ensembles used in this study.
We also document the trajectories used for $m_{\rm res}$ measurements,
using every tenth trajectory in the given range.
The trajectory length was $\tau=0.5$ units of MD time.
All ensembles used three degenerate DWF flavors of mass $m = 0.04$,
and $M_5 = 1.8$. }

\begin{tabular}{|\cs c \cs| \cs c \cs | \cs c \cs | \cs c \cs |}
\hline
Gauge Action & $\beta$ & Algorithm  & Trajectories\\
\hline
DBW2  & 0.72  & R   &  800-1500\\ 
DBW2  & 0.764 & R   &  800-1500\\
DBW2  & 0.78  & R   &  1000-1600\\
DBW2  & 0.80  & R   &  900-1800\\
DBW2  & 0.88  & R   &  1000-3000\\
Iwasaki& 2.13 & RHMC&  1500-2400\\
Iwasaki& 2.2  & RHMC&  800-1500\\
Iwasaki& 2.3  & RHMC&  800-1500\\
\hline
\end{tabular}
\end{table}

\clearpage

\begin{table}[hbt]
\caption{
\label{TabMresData}
Correlation functions generated for
valence quarks with different extents of the fifth dimension
on each of our $L_s=8$ dynamical ensembles. 
The valence and sea quark masses were $m_{ud} = m_s = 0.04$.
We denote point source correlators as ``P'', wall
source correlators as ``W'', and exponentially smeared correlators as ``S''.
}
\begin{tabular}{|\cs c \cs |\cs ccccccc \cs|}
\hline
Dataset & $L_s=8$ &$L_s=12$ &$L_s=16$ & $L_s=20$ &$L_s=24$ &$L_s=28$ &$L_s=32$ \\
\hline
Iwasaki $\beta=2.13$& S   & S & S & S & S & - & S\\ 
Iwasaki $\beta=2.2$ & W,P & P & P & P & P & - & P\\  
Iwasaki $\beta=2.3$ & W,P & P & P & P & P & - & P\\  
DBW2 $\beta=0.72$  & W   & W & W & - & W & - & W\\
DBW2 $\beta=0.764$ & W   & P & P & P & P & P & P\\  
DBW2 $\beta=0.78$  & W   & W,P & W,P & W,P & W,P & P & P\\  
DBW2 $\beta=0.80$  & W   & P & P & P & P & P & P\\  
DBW2 $\beta=0.88$  & P & W,P & W,P & W,P & W,P & - & W,P\\  
\hline
\end{tabular}
\end{table}

\clearpage

\begin{sidewaystable}
\centering
\caption{
\label{TabMresDBW2}
Fitted residual mass values for valence quarks with different extents of the fifth dimension for the 
DBW2 gauge action 
on each of our $L_s=8$ dynamical ensembles. The valence and sea quark masses were $m_{ud} = m_s = 0.04$.
}
\begin{tabular}{|\cs c\cs|\cs ccccccc\cs|}
\hline
Dataset & $L_s=8$ &$L_s=12$ &$L_s=16$ & $L_s=20$ 
        & $L_s=24$ &$L_s=28$ &$L_s=32$ \\
\hline
$\beta=0.72$  & $1.13(1)\times10^{-2}$
              & $ 4.4(1)\times10^{-3} $
              & $ 2.2(1)\times10^{-3} $
              &  - 
              & $1.04(5) \times10^{-3}$
              & - 
              & $7.0(5) \times10^{-4}$\\

$\beta=0.764$ & $5.46(4)\times10^{-3}$  
              & $1.53(3) \times10^{-3}$ 
              & $5.9(2)\times10^{-4}$   
              & $3.0(2)\times10^{-4}$   
              & $2.0(2)\times10^{-4}$   
              & $1.5(2)\times10^{-4}$   
              & $1.2(2)\times10^{-4}$\\ 

$\beta=0.78$  & $4.28(3) \times10^{-3}$ 
              & $1.09(2)\times10^{-3}$  
              & $3.7(2)\times10^{-4}$   
              & $1.7(1)\times10^{-4}$   
              & $1.0(2)\times10^{-4}$   
              & $7(1) \times10^{-5} $   
              & $6(1) \times10^{-5} $\\ 

$\beta=0.80$  & $3.39(2)\times10^{-3}$  
              & $8.4(2)  \times10^{-4}$ 
              & $2.7(1) \times10^{-4}$  
              & $1.2(1)\times10^{-4}$   
              & $6.6(8)\times10^{-5}$   
              & $4.6(7)\times10^{-5}$   
              & $3.6(7)\times10^{-5}$\\ 

$\beta=0.88$  & $1.441(4)\times10^{-3}$ 
              & $2.40(2)\times10^{-4}$  
              & $4.87(6)\times10^{-5}$  
              & $1.14(3)\times10^{-5}$  
              & $3.0(1)\times10^{-6}$   
              & -
              & $3.3(4)\times10^{-7}$\\    
\hline
\end{tabular}
\end{sidewaystable}

\begin{sidewaystable}
\centering
\caption{
\label{TabMresIwasaki}
Fitted residual mass values for valence quarks with different extents of the fifth dimension for the Iwasaki 
gauge action on each of our $L_s=8$ dynamical ensembles. The valence and sea quark masses were $m_{ud} = m_s = 0.04$.
}
\begin{tabular}{|\cs c \cs| \cs ccccccc \cs|}
\hline
Dataset & $L_s=8$ &$L_s=12$ &$L_s=16$ & $L_s=20$ &
        $L_s=24$  &$L_s=32$ \\
\hline
$\beta=2.13$  & $1.18(1)\times10^{-2}$ 
              & $4.53(5)\times10^{-3}$ 
              & $2.25(5)\times10^{-3}$ 
              & $1.38(4)\times10^{-3}$ 
              & $9.8(4)\times10^{-4}$  
              & $6.3(3)\times10^{-4}$ \\ 
         
$\beta=2.2$ & $7.19(6) \times10^{-3}$ 
            & $2.25(4) \times10^{-3}$ 
            & $9.0(3) \times10^{-4}$  
            & $4.6(2) \times10^{-4}$  
            & $2.9(2) \times10^{-4}$  
            & $1.7(2) \times10^{-4}$\\

$\beta=2.3$ & $3.90(2)\times10^{-3}$ 
            & $9.2(1) \times10^{-4}$ 
            & $2.65(6)\times10^{-4}$ 
            & $9.5(5) \times10^{-5}$ 
            & $4.2(4) \times10^{-5}$ 
            & $2.2(4) \times10^{-5}$ 
\end{tabular}
\end{sidewaystable}

\clearpage

\begin{table}[hbt]
\caption{
\label{TabFitMresIwasaki}
Fit parameters to the dependence of $m_{\rm res}$ on $L_s$ 
for the Iwasaki and DBW2 gauge actions.
We also give the lattice spacing obtained from the static potential 
\cite{Hashimoto:2005re,Antonio:2006px} on the
dynamical background for $m_{ud} = m_s = 0.04$
without chiral extrapolation. This differs from the lattice spacing defined in the
chiral limit presented in \cite{Antonio:2006px}, and we only provide the lattice spacings
here for illustration.
}
\begin{tabular}{|\cs c\cs|\cs c\cs|\cs cccc\cs \cs|}
\hline
Action & $\beta$ & $c_1$ & $\lambda_c$ & $c_2$ & $a^{-1}$(GeV)\\
\hline
Iwasaki & 2.13 & 0.344(6) & 0.191(2) & 0.0198(3) & 1.62(2) \\
Iwasaki & 2.2  & 0.301(4) & 0.219(2) & 0.0054(1) & 1.89(2) \\
Iwasaki & 2.3  & 0.262(3) & 0.268(2) & 0.0006(1) & 2.26(7)\\
\hline
DBW2 & 0.72    & 0.342(10)& 0.201(4) & 0.0220(4)     & 1.40(4)\\
DBW2 & 0.764   & 0.298(6) & 0.252(3) & 0.0040(1)     & 1.74(3)\\
DBW2 & 0.78    & 0.267(5) & 0.264(2) & 0.00188(7)    & 1.82(2)\\
DBW2 & 0.80    & 0.216(5) & 0.265(3) & 0.00120(5)    & 1.98(4)\\
DBW2 & 0.88    & 0.171(7) & 0.338(5) & 1.0(5)$\times10^{-5}$ & - \\
\hline
\end{tabular}
\end{table}


\clearpage

\clearpage

\clearpage

\clearpage


\clearpage

\clearpage

\end{document}